\documentclass[%
 reprint,
 amsmath,amssymb,
 aps,
]{revtex4-2}

\usepackage{graphicx}
\usepackage{dcolumn}
\usepackage{bm}

\begin{document}

\preprint{APS/123-QED}

\title{Controlling the shape and topology of two-component colloidal membranes}


\author{Ayantika Khanra$^1$, Leroy L. Jia$^2$, Noah P. Mitchell$^{3,4}$, Andrew Balchunas$^5$, Robert A. Pelcovits$^6$, Thomas R. Powers$^{7,6}$, Zvonimir Dogic$^{4,5,8}$, Prerna Sharma$^{1,9}$}
 \email{Email for correspondence: prerna@iisc.ac.in}

\affiliation{$^1$ Department of Physics, Indian Institute of Science, Bangalore 560012, India}%
\affiliation{$^2$ Center for Computational Biology, Flatiron Institute, New York, NY 10010, USA}
\affiliation{$^3$ Kavli Institute for Theoretical Physics, University of California, Santa Barbara, CA 93106, USA}
\affiliation{$^4$ Physics Department, University of California, Santa Barbara, CA 93106, USA}
\affiliation{$^5$ Martin A. Fisher School of  Physics, Brandeis University, Waltham, MA 02454, USA}
\affiliation{$^6$ Brown Theoretical Physics Center and Department of Physics, Brown University, Providence, RI 02912, USA}
\affiliation{$^7$ Center for Fluid Mechanics and School of Engineering, Brown University, Providence, RI 02912, USA}
\affiliation{$^8$ Biomolecular Science and  Engineering Department, University of California, Santa Barbara, CA 93106, USA}
\affiliation{$^9$ Centre for Biosystems Science and Engineering, Indian Institute of Science, Bangalore 560012, India}

\date{\today}

\begin{abstract}
Changes in the geometry and topology of self-assembled membranes underlie diverse processes across cellular biology and engineering. Similar to lipid bilayers, monolayer colloidal membranes have in-plane fluid-like dynamics and out-of-plane bending elasticity. Their open edges and micron length scale provide a tractable system to study the equilibrium energetics and dynamic pathways of membrane assembly and reconfiguration. 
Here, we find that doping colloidal membranes with short miscible rods transforms disk-shaped membranes into saddle-shaped surfaces with complex edge structures. The saddle-shaped membranes are well-approximated by Enneper's minimal surfaces. Theoretical modeling demonstrates that their formation is driven by increasing positive Gaussian modulus, which in turn is controlled by the fraction of short rods. Further coalescence of saddle-shaped surfaces leads to diverse topologically distinct structures, including catenoids, tri-noids, four-noids, and higher order structures. At long time scales, we observe the formation of a system-spanning, sponge-like phase. The unique features of colloidal membranes reveal the topological transformations that accompany coalescence pathways in real time. We enhance the functionality of these membranes by making their shape responsive to external stimuli. Our results demonstrate a novel pathway towards control of thin elastic sheets' shape and topology --- a pathway driven by the emergent elasticity induced by compositional heterogeneity.
\end{abstract}

\maketitle

Thin sheets can assume diverse geometrical and topological shapes and structures, which permeate the natural world across length scales. At the cellular level, nanometer-thick fluid-like lipid membranes can seamlessly transition between distinct topological structures, a unique feature that is essential for endo- and exocytosis, viral infection, as well as transport of nutrients and signaling molecules~\cite{ConfMembVesicles,Almsherqi2006,Michalet666,Longley1983,McMahon2005,kozlov1,snapp2003formation,bussi2019fundamental,harrison2008viral,zimmerberg2006proteins,bykov2017structure}. At organismal scales, microns-thin cellular sheets can transform into complex tubular, coiled, and branched structures that underlie morphogenesis of flowers, visceral organs, and the nervous system~\cite{lewicka2021geometry,savin2011growth,van2017growth,mitchell2021visceral,karzbrun2021human,metzger2008branching}. Designing responsive synthetic materials that can assume the above-described 3D shapes and topologies observed in biology remains a challenge. So far, work has primarily focused on macroscale stimuli-responsive solid-like elastic sheets that have a finite in-plane shear modulus~\cite{PhysRevLett.90.074302, Kim1201,klein2007shaping,Huang650}. Such materials allow one to engineer lateral stress patterns that yield targeted 3D architectures. However, solid elastic sheets cannot easily fuse into non-trivial topologies. In comparison, nanometer-sized fluid lipid bilayers seamlessly transform between various topologically complex surfaces, but these transitions occur on time and length scales that preclude real time observation. The limitations of both the nanoscale fluid bilayers and the macroscale solid elastic sheets reveal a need for an experimental platform to study formation of topologically complex surfaces.

Motivated by such considerations, we study colloidal membranes, which are micron-thick fluid-like monolayers of aligned rod-like particles. They share many properties with lipid bilayers, but at larger length scales and slower time scales~\cite{barry2010entropy}. Introducing a critical fraction of miscible shorter rods destabilizes the flat state, leading to the formation of diverse geometrically and topologically complex surfaces. The length scale of the colloidal membranes enables visualization of 3D pathways by which open flat sheets transform into topologically non-trivial structures. A continuum model describes the membrane shape transitions by balancing the edge energy, which favors flat disk-shaped membranes, with the Gaussian curvature modulus which favors saddle-shaped structures.  

\section*{Results}

\subsection*{Assembly of colloidal membranes}
Colloidal membranes are one-rod-length thick fluid monolayers composed of aligned rods that self-assemble in the presence of depleting polymers~\cite{barry2010entropy,yang2012self,Balchunas}. They form by a robust assembly pathway that does not require chemical heterogeneity such as an amphiphilic nature of the building blocks, but rather relies on the anisotropy of the building blocks. Entropic depletion interactions favor alignment of the rods along their long axes, which minimizes volume excluded to the polymer coils~\cite{Asakura_Interaction_2_bodies}. The attraction strength is determined both by the length of the rods and the concentration of the depleting polymer. Over an intermediate range of depletant concentrations, such tunable attractions assemble membrane-like materials, in which there is a complete phase separation between the rod-rich membrane and a polymer suspension that envelops the membrane. The density of the rods within a membrane is determined by the osmotic pressure that is exerted by the immiscible polymers. Similar to lipid bilayers, the out-of-plane deformations of colloidal membranes are described by the Helfrich Hamiltonian~\cite{helfrich1973elastic,gibaud2017achiral,jia2017chiral}. However, unlike lipid bilayers which assemble into edgeless vesicles, colloidal membranes usually assume flat 2D disk-like shapes. Furthermore, being assembled from one-micron long particles, colloidal membranes allow for visualization of various in-plane structures and dynamical pathways that are not easily studied with nanometer sized lipid bilayers~\cite{gibaud2012reconfigurable,barry2008direct,sharma2014hierarchical,miller2019conformational,miller2020all}. 
\begin{figure*}[hb!]
\centering
\includegraphics[width=\linewidth ]{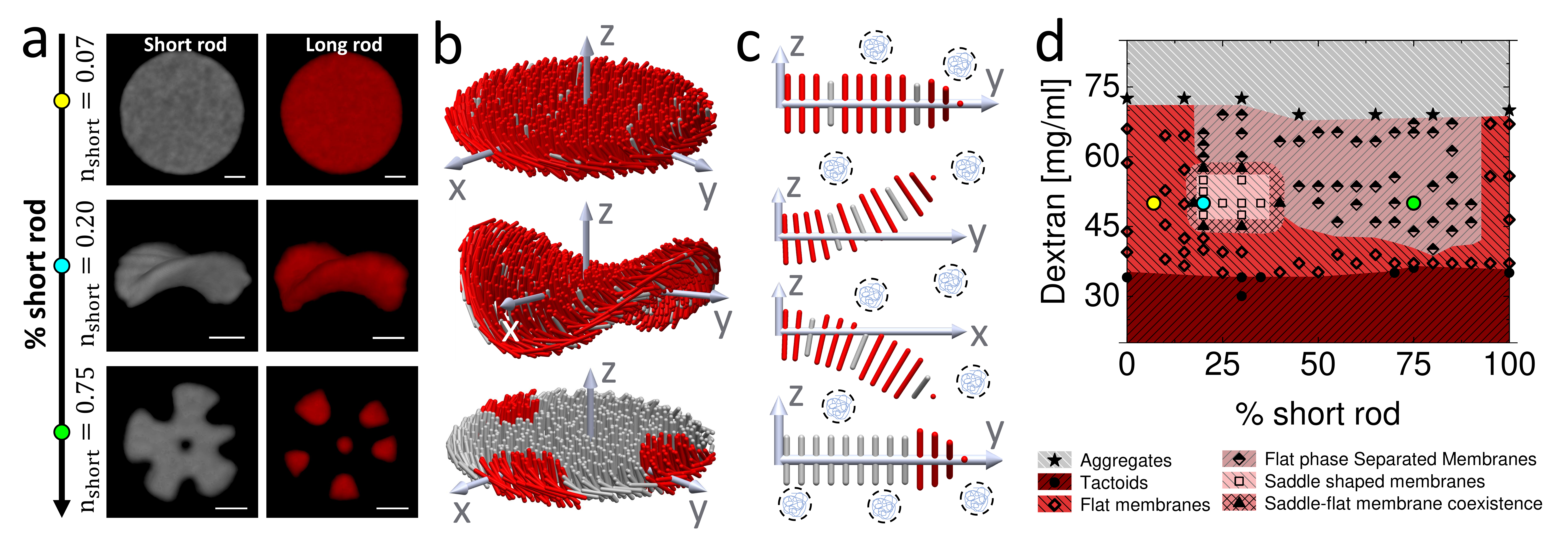}
\caption{\textbf{Phase diagram of binary colloidal membranes.} \textbf{(a)} Two-component colloidal membranes composed of long (red) and short (gray) rods. Top to bottom: Increasing the number fraction of short rods, $n_\textrm{short}$, causes uniformly mixed flat membranes to transform into saddle shapes. At higher $n_\textrm{short}$, phase separation between the two types of rods occurs, and the membranes revert to flat shapes. Dextran concentration, 50\,mg/ml; scale bar, 2 \textmu m. \textbf{(b-c)} Schematics of the rod positions and orientations within membranes at different $n_\textrm{short}$. \textbf{(d)} Phase diagram of long and short rod mixtures: Disordered aggregates form above $\sim$ 68\,mg/ml Dextran concentration (filled stars) and tactoids form below $\sim$ 35 mg/ml Dextran concentration (filled circles). At intermediate Dextran concentrations, the following types of colloidal membranes are observed: flat and phase-separated (half-filled diamonds), flat and uniformly mixed (open diamonds), and membranes with negative Gaussian curvatures (open squares). Filled triangles correspond to coexistence of flat membranes and saddles. Yellow, cyan, and green filled circles correspond to the assembly conditions of the membranes shown in panel a.}
\label{fig: PhaseDiag_Schematics}
\end{figure*}

\subsection*{Short rods destabilize flat colloidal membranes}
We studied binary colloidal membranes assembled from two chiral filamentous viruses with the same handedness, 1200\,nm-long M13KO7 and 880\,nm-long M13-wt~\cite{dogic2001development,grelet2003origin,barry2009model}. In the presence of the non-adsorbing polymer Dextran, both virus types co-assembled into 2D colloidal membranes. We first changed the number fraction of short rods, $n_\textrm{short}$, within the membrane while keeping the Dextran concentration fixed. The common chirality of the rods increased the miscibility of the two species, when compared to rods of opposite chirality~\cite{sharma2014hierarchical,siavashpouri2019structure}. At low number fractions of short M13-wt rods ($n_\textrm{short}<0.15$) we observed assembly of flat colloidal membranes. Labeling both rod types revealed uniformly mixed membranes  (Fig.~\ref{fig: PhaseDiag_Schematics}(a-c), top row). At intermediate volume fractions ($0.2<n_\textrm{short}<0.35$) the membranes assumed 3D saddle-like surfaces with negative Gaussian curvature (Fig.~\ref{fig: PhaseDiag_Schematics}(a-c), middle row). Labeling both virus types revealed that saddle-membranes remained uniformly mixed for $n_\textrm{short}<0.25$. Beyond this fraction, we observed that short rods started phase separating at the membrane's edge, but the interior remained uniformly mixed. Increasing the number fraction of short rods even further ($n_\textrm{short}>0.35$) yielded another transition from saddle surfaces back to flat membranes. This was accompanied by an in-plane phase separation into two phases that were respectively enriched in the long and short rods (Fig.~\ref{fig: PhaseDiag_Schematics}(a-c), bottom row). 

We mapped the phase diagram as a function of $n_\textrm{short}$ and  Dextran concentrations (Fig.~\ref{fig: PhaseDiag_Schematics}d). Liquid crystalline tactoids and disordered smectic-like stacks formed at low and high Dextran concentrations, respectively~\cite{dogic2003surface}. Colloidal membranes formed at intermediate Dextran concentrations~\cite{barry2010entropy}. For a fixed intermediate Dextran concentration, we found that increasing the fraction of short rods, first led to flat and miscible membranes, then saddle-shaped miscible membranes, and finally phase-separated flat membranes. At the transition points, flat and saddle-shaped membranes coexisted.

\begin{figure*}[ht!]
\centering
\includegraphics[width=\linewidth ]{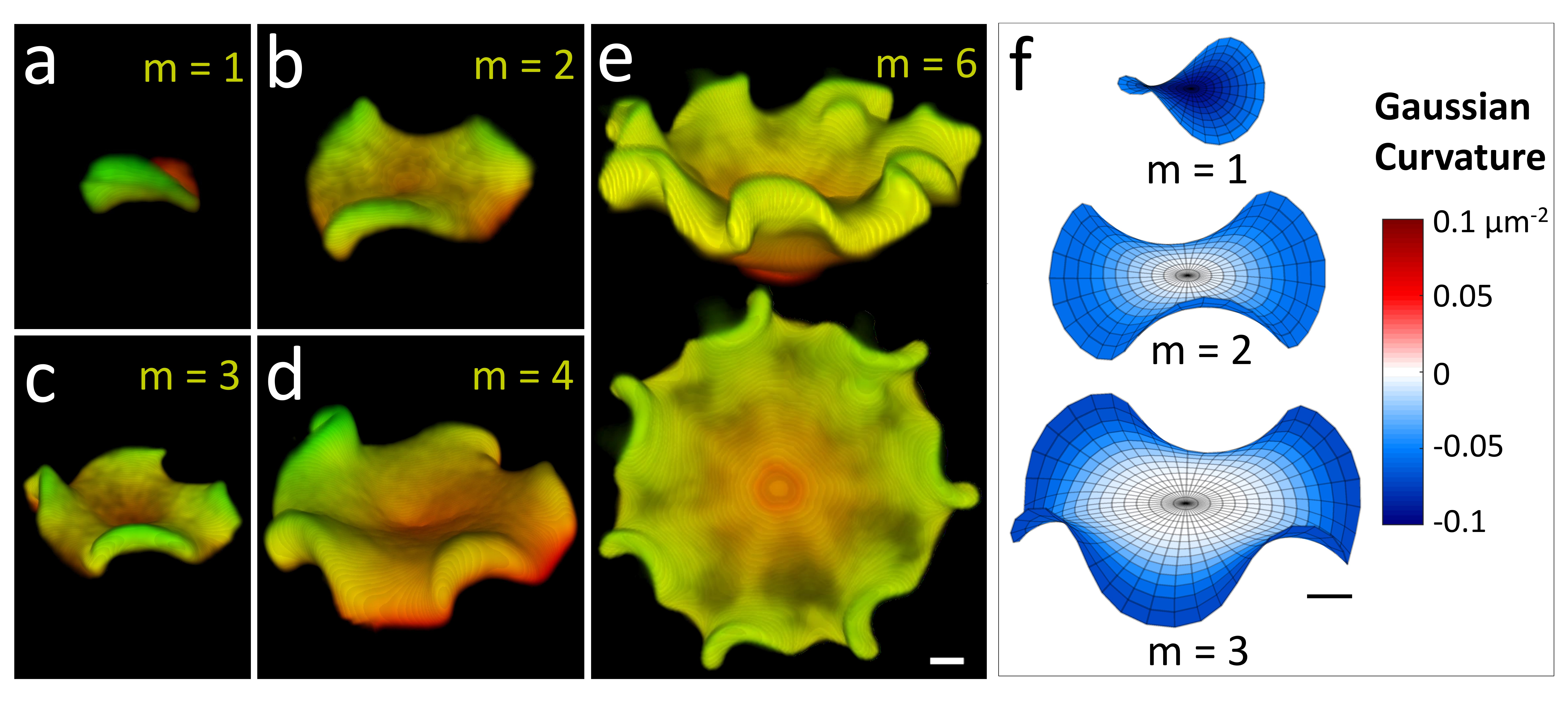}
\caption{\textbf{Saddle-shaped colloidal membranes mimic Enneper surfaces.} \textbf{(a-d)} Confocal images of saddle-shaped colloidal membranes with increasing order number $m$ = 1, 2, 3 and 4. \textbf{(e)} 3D rendered images of deconvolved wide field z-stack of a $m$ = 6 order saddle surface. Scale bar, 2 \textmu m. Dextran concentration, 50\,mg/ml and $n_\textrm{short}=0.2$. \textbf{(f)} Enneper surfaces of orders $m$=1, 2 and 3. Color indicates the Gaussian curvature. The area increases with the surface order $m$, a trend observed in the experiments. Scale bar, 2 \textmu m.}
\label{fig: Higher_order_saddles}
\end{figure*}

Next, we studied the coalescence kinetics of saddle-shaped membranes. Observing the sample over time elucidates how increasing the membrane area affects its 3D shape. Immediately upon preparation, all rods condensed into small colloidal membranes, with lateral size of a few microns. Over time, their size increased as the membranes laterally coalesced with each other. Initially, the samples contained colloidal membranes that were mostly saddle-shaped (Fig.~\ref{fig: Higher_order_saddles}a-e). These saddle-like surfaces can be classified according to their order, which characterizes the number of minima and maxima encountered as one moves along the membrane perimeter. Intriguingly, the saddle-shaped membranes exhibit asymmetric distortions along their perimeter, which might be due to the chiral nature of the constituent rods. Even for flat membranes, chirality induces asymmetric rippled edge fluctuations~\cite{jia2017chiral}. We observed up to sixth order saddle-membranes  (Fig.~\ref{fig: Higher_order_saddles}e and SI Appendix, Fig.~S1).

\subsection*{Saddle-shaped membranes are Enneper-like minimal surfaces}
Saddle-shaped colloidal membranes (Fig.~\ref{fig: saddles_th_exp}a) have negative Gaussian curvature and small mean curvature. To quantify their shape, we extracted the membrane midplane from 3D confocal images and computed the spatial maps of both their Gaussian curvature $K$ and mean curvature $H$ (SI Appendix, SI Text). The median value of the Gaussian curvature was negative ($\langle K\rangle=-0.063\,\mu\mathrm{m}^{-2}$), while the median mean curvature was much smaller when compared to the square root of the Gaussian curvature ($\langle H\rangle=0.010\,\mu\mathrm{m}^{-1}\ll\sqrt{|\langle K\rangle|}$). Both curvatures had small scale heterogeneity due to measurement noise (Fig.~\ref{fig: saddles_th_exp}b,c). These results suggest that saddle-membranes are minimal-like surfaces.

Inspection of different minimal surface families revealed a resemblance between saddle-membranes and Enneper minimal surfaces (Fig.~\ref{fig: Higher_order_saddles}f)~\cite{FomenkoTuzhilin1991}. The Enneper minimal surfaces have the parametrization
\begin{eqnarray}
\frac{x}{R}&=&r\cos\phi-\frac{r^{2m+1}}{2m+1}\cos[(2m+1)\phi]\label{xeqn}\\
\frac{y}{R}&=&r\sin\phi+\frac{r^{2m+1}}{2m+1}\sin[(2m+1)\phi]\label{yeqn}\\
\frac{z}{R}&=& \frac{2r^{m+1}}{m+1}\cos[(m+1)\phi],\label{zeqn}
\end{eqnarray}
where $r$ is a dimensionless radial coordinate, $\phi$ is the angular coordinate, $m$ is the order, and $R$ is a positive parameter~\cite{FomenkoTuzhilin1991}. We take the surface area to be $A=\pi R_0^2$. When $R$ is much greater than $R_0$, the saddle is gently curved and has a disk-like shape. As $R$ decreases toward $R_0$, the surface curvature increases, and the surface self-intersects if $R$ becomes sufficiently small. By adjusting the $R$ parameter, we found Enneper surfaces that well approximate the shapes of representative saddle-membranes, with some discrepancies being observed close to the membrane's edge (Fig.~\ref{fig: saddles_th_exp}d). Furthermore, the edge of the experimental surface did not extend the same distance in all directions. For instance, the surface extended much further along one diagonal when compared to the other one (lines ii and~iv in Fig.~\ref{fig: saddles_th_exp}d). The fits for the higher order experimental surfaces were also less accurate due to the appearance of a central bulge and the chiral structure of the edge (Fig.~\ref{fig: Higher_order_saddles}e, top). 

\subsection*{Theoretically determining the shape of the membrane edge}
We introduce a theoretical model to determine the shape of the membrane edge, which is given by $r_\mathrm{edge}(\phi)$.  Since the tilt of the rods within a saddle-membrane is small everywhere except near the edge, we use a continuum model with an effective edge energy to account for the liquid crystalline degrees of freedom close to the edge. This model accounts for the membrane composition implicitly, through the assumption that changing the fraction of long and short rods primarily affects the values of the elastic moduli. 

The conformation of a colloidal membrane is described by an energy that accounts for a resistance to bending~\cite{canham1970minimum,helfrich1973elastic}, a fixed area, a free edge with an edge tension, a resistance to edge bending, and a preference for a twist due to the chiral constituents~\cite{jia2017chiral,balchunas2020force}:
\begin{eqnarray}
E&=&\frac{\kappa}{2}\int\mathrm{d}A (2H)^2+\bar{\kappa}\int\mathrm{d}A K+\mu\int\mathrm{d}A\nonumber\\
&+&\gamma\int\mathrm{d}l+\frac{B}{2}\int\mathrm{d}l k^2+\frac{B^\prime}{2}\int\mathrm{d}l(\tau_\mathrm{g}-\tau^*_\mathrm{g})^2,\label{CHEnergy}
\end{eqnarray}
where $\kappa$ is the bending modulus, $\bar{\kappa}$ is the Gaussian curvature modulus, $\mathrm{d}A$ is the element of area of the membrane mid-surface, $\mu$ is a Lagrange multiplier enforcing area conservation, $\gamma$ is the edge tension, $\mathrm{d}l$ is the element of arclength of the edge, $k$ is the curvature of the edge, $\tau_\mathrm{g}$ is the geodesic torsion~\cite{kleman}, $\tau^*_\mathrm{g}$ is the spontaneous geodesic torsion (proportional to the desired rate of twist), and $B$ and $B^\prime$ are the elastic moduli associated with the edge. It is convenient to work in terms of the chirality modulus $c^*= -B'\tau_\mathrm{g}^*$. 

We can independently estimate the magnitude of the phenomenological parameters in Eq.~\ref{CHEnergy}. Measurements of the fluctuations of the edge of a disk-like membrane ($n_{\mathrm{short}}=0.1$) reveal that  $B\approx150~k_BT\,$\textmu m and $\gamma_\mathrm{eff}\approx620~k_BT/$\textmu m (SI Appendix, Fig.~S2)~\cite{gibaud2012reconfigurable}. There is a chiral contribution to the line tension $\gamma_{\mathrm{c}} =  (c^*)^2/(2B')$, so that the effective total line tension is $\gamma_\mathrm{eff}=\gamma + \gamma_{\mathrm{c}}$ \cite{gibaud2012reconfigurable}. We also assume that  $B^\prime=B$.   Furthermore, colloidal membranes' area compressibility implies a large value of the bending modulus, $\kappa\approx 15,000\,k_BT$~\cite{Balchunas_etal2019}. Membranes with positive Gaussian modulus decrease their energy by adopting negative Gaussian curvature, and the Gaussian moduli of both single component fd-wt and mixed fd-wt/fd-Y21M membranes have a magnitude of $\sim 200\,k_BT$ and are positive~\cite{gibaud2017achiral,jia2017chiral}. Thus we expect $\kappa\gg\bar{\kappa}$, which provides an additional rationale for modeling saddles as Enneper minimal surfaces (Fig.~\ref{fig: Higher_order_saddles}f). 

Assuming that the non-flat membranes are described by $m$th-order Enneper surfaces, we solve the in-plane force balance equation for the edge to determine the boundary contour, $r_{\mathrm{edge}}(\phi)$, subject to the constraints of the fixed area and periodicity in $\phi$ (SI Appendix, Eq. 3). The minimal surface assumption ($\bar\kappa \ll \kappa$) implies that the other equilibrium conditions for out-of-plane force balance (SI Appendix, Eq. 2) and moment balance~(SI Appendix, Eq. 4) at the membrane's edge are trivially satisfied, since the terms proportional to $\kappa H$ or $\kappa \boldsymbol{\nabla}H$ can take whatever values are necessary to satisfy these equations when $\kappa\rightarrow\infty$ and $H\rightarrow0$. 

\begin{figure*}[ht!]
\centering
\includegraphics[width=\linewidth ]{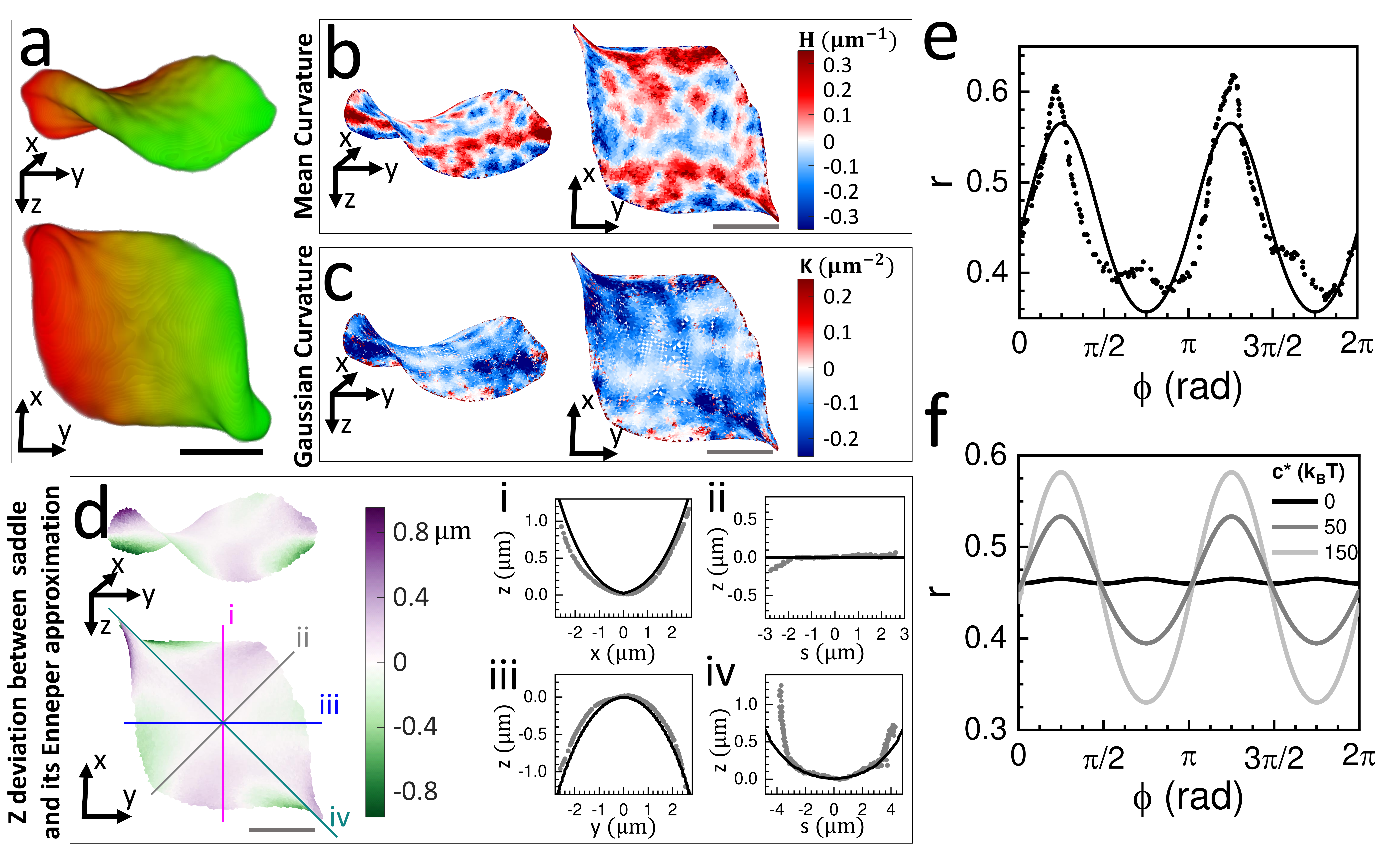}
\caption{\textbf{Saddle-shaped membranes are nearly Enneper minimal surfaces.} \textbf{(a)} Top and side view of a saddle membrane, rendered from confocal z-stacks. \textbf{(b-c)} Top and side views of the mid-surface of the saddle membrane. Color indicates the local mean and Gaussian curvature in (b) and (c), respectively. The membranes have predominantly negative Gaussian curvature and nearly zero mean curvature. \textbf{(d)} Top and side view comparison between an experimental mid-surface and the numerical model. Color indicates deviation between the best-fit Enneper surface (size parameter, $R=6.47$ \textmu m) and the membrane along the $z$-axis. Lines i-iv show height profiles of the mid-surface (gray dots) and the best-fit Enneper surface (black curves). Scale bars, 2 \textmu m. \textbf{(e)} $r$-$\phi$ parameterization of the edge of the saddle shown in (b). Black dots are experimental measurements and the black curve is a theoretical prediction in which the Gaussian and chirality moduli are set to $1000~k_BT$ and $100~k_BT$, respectively. \textbf{(f)} In theoretical predictions, increasing the chirality modulus elongates the saddle in diametrically opposite directions, resulting in two peaks in $r$-$\phi$ plot of the saddle's edge.}
\label{fig: saddles_th_exp}
\end{figure*}

The numerical model correctly captures the $(m+1)$-fold dihedral symmetry of the Enneper surface. Generically, we find that a one-mode approximation $r_\mathrm{edge}(\phi) \approx r_0 + r_1 \cos[(m+1)(\phi-\phi_0)]$, for constants $r_0$, $r_1$ and $\phi_0$, is sufficient to describe the solutions to Eq. 3 (SI Appendix). Notably, the amplitude $r_1$ depends weakly on $\bar\kappa$ but strongly on $c^*$ (Fig.~\ref{fig: saddles_th_exp}f). However, increasing $c^*$ or $\bar\kappa$ decreases the value of $R$ that minimizes the energy $E$, since these changes cause the surface to prefer additional curvature.  
 
We determine $\bar\kappa$ and $c^*$ by fitting the edge profiles of the saddle surfaces. During the fitting process, $\gamma$ and $B$ are fixed at the values mentioned above, and $R$ is fixed at the value obtained from fits of the mid-surface. For the saddle shown in Fig.~\ref{fig: saddles_th_exp}a-d ($R=6.47$ \textmu m), $\bar\kappa \approx 1000$ $k_BT$ and $c^*\approx 100$ $k_BT$ (Fig.~\ref{fig: saddles_th_exp}e). Fitting of the edge profiles of other saddles of various orders yields similar results. We note that the magnitude of the chiral modulus is comparable to previous estimates~\cite{balchunas2020force,jia2017chiral}, while $\bar\kappa$ is nearly an order of magnitude greater than the value for single component membranes. Also, recall that the Gauss-Bonnet theorem shows that the Gaussian curvature energy is effectively an edge energy, despite the fact that it is written as an integral over the entire surface in Eq.~\ref{CHEnergy}. 

\subsection*{Continuum model estimates the stability of  saddle-shaped membranes} 
\begin{figure}
\centering
\includegraphics[width=\linewidth ]{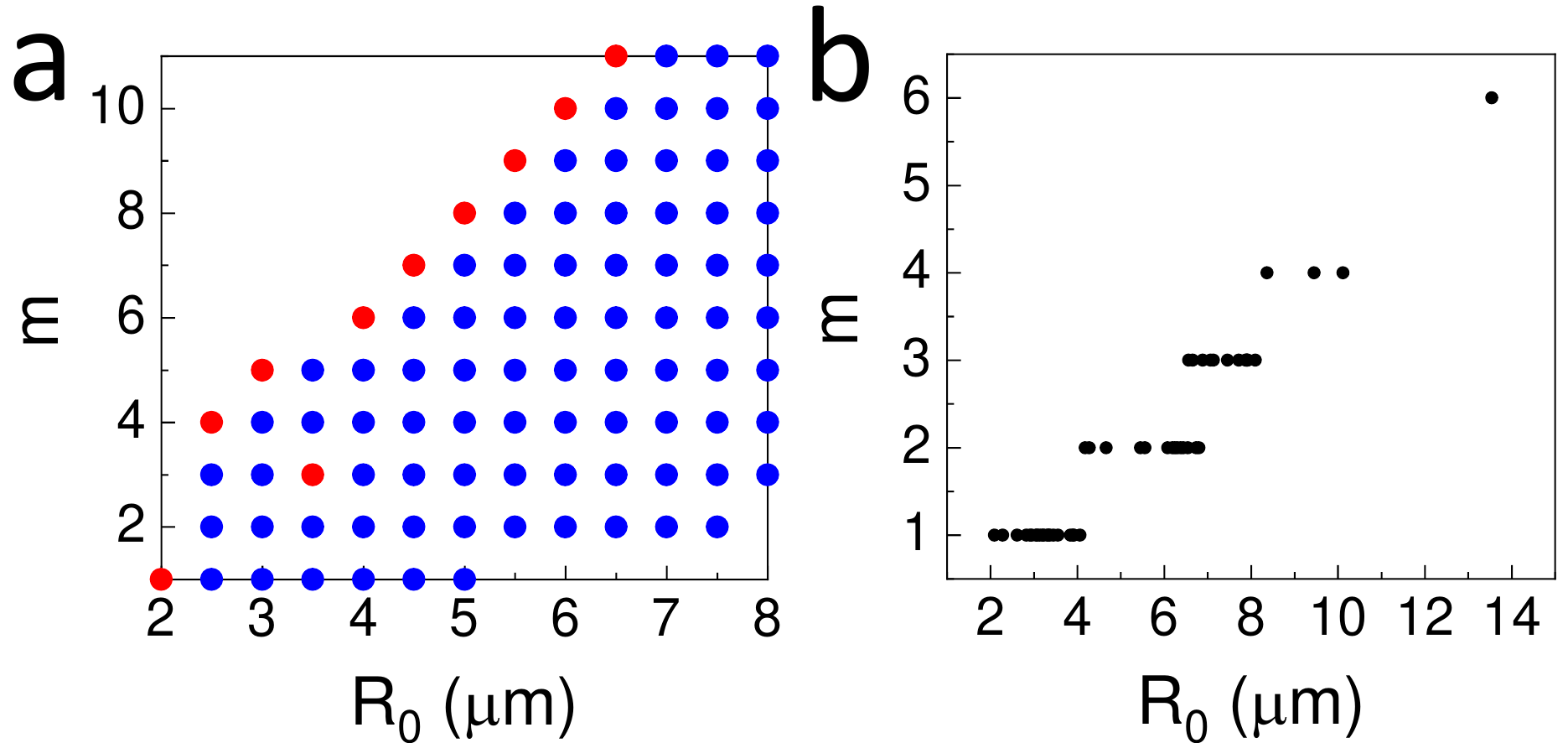}
\caption{\textbf{Saddle order increases with increase in surface area.} \textbf{(a)}
The values of $m$ for which there exist Enneper surfaces that are stable and lower energy than the disk of radius $R_0 = \sqrt{A/\pi}$. Blue dots indicate the existence of such a saddle; red dots indicate the stable saddle of lowest energy; for most areas, this happens at the maximum allowable $m$. Energies were computed for a membrane with $\gamma = 622~k_BT/$\textmu m, $B = B' = 155~k_BT\,$\textmu m, $c^* = 100~k_BT$, and $\bar\kappa = 1000~k_BT$. As area increases, the lower order saddles become less energetically favorable compared to the disk while the higher order saddles become more energetically favorable. Saddles of order up to and including $m=11$ were compared in this plot. \textbf{(b)} Ranges of experimentally observed saddle sizes and corresponding saddle orders. Membrane z-stacks were taken from a combination of confocal and deconvolution microscopy; the smallest membranes ($R_0<2$ \textmu m) were not processed. While the theory tends to overestimate $m$, the general trend of $m$ linearly increasing as $R_0$ increases is clearly visible.}
\label{fig:Multicrit}
\end{figure}

Using our continuum model, we estimate the stability of Enneper-like surfaces. Our focus is on the transition between flat disks and Enneper surfaces, and between Enneper surfaces of different orders. We use the above-described parameter values to minimize Eq.~\ref{CHEnergy}. For a given area, we calculate the energies of the eleven lowest order saddles using the numerical procedure outlined above as functions of $R$, which is determined numerically through minimization. While driving $R$ to zero almost always results in an absolute minimum, local minima where $\partial E/\partial R=0$ and $\partial^2 E/\partial R^2  >0$ also occur for certain parameter combinations. Figure~\ref{fig:Multicrit}a shows a range of areas $R_0 = \sqrt{A/\pi}$ and the values of $m$ for which  Enneper surfaces have lower energy than a flat disk of equivalent area. The effects of a larger or smaller $\bar\kappa$ were also explored (SI Appendix, SI Text and Fig.~S3).

For the smallest membrane sizes, only the lowest order saddles are stable, since higher order Enneper surfaces have more Gaussian curvature. As $R_0$ increases, the theoretical range of allowable saddles broadens, with the energy minimizing saddle having the highest allowable $m$. There is one exception at $R_0 =3.5$ \textmu m which is possibly due to the existence of multiple solution branches (Supplementary Information). At the largest areas, the lowest order saddles are no longer favorable compared to the disk since the edge cost at small $m$ is larger than the Gaussian curvature energy gained by curving it into a saddle. None of the local minima exhibit self-intersection. The emergence of new energy minimizing saddles as area is increased appears to be a first order transition.

This phase diagram is in qualitative agreement with the experimental result that low order saddles are observed for small area membranes and increasing the area leads to higher order saddles, with a dependence that appears almost linear (Figure~\ref{fig:Multicrit}b). While the theory tends to overestimate the order of saddles at higher areas, it also does not account for external barriers to formation of non-Enneper shapes.

\subsection*{Topological transitions}
\begin{figure*}[ht!]
\centering
\includegraphics[width=\linewidth ]{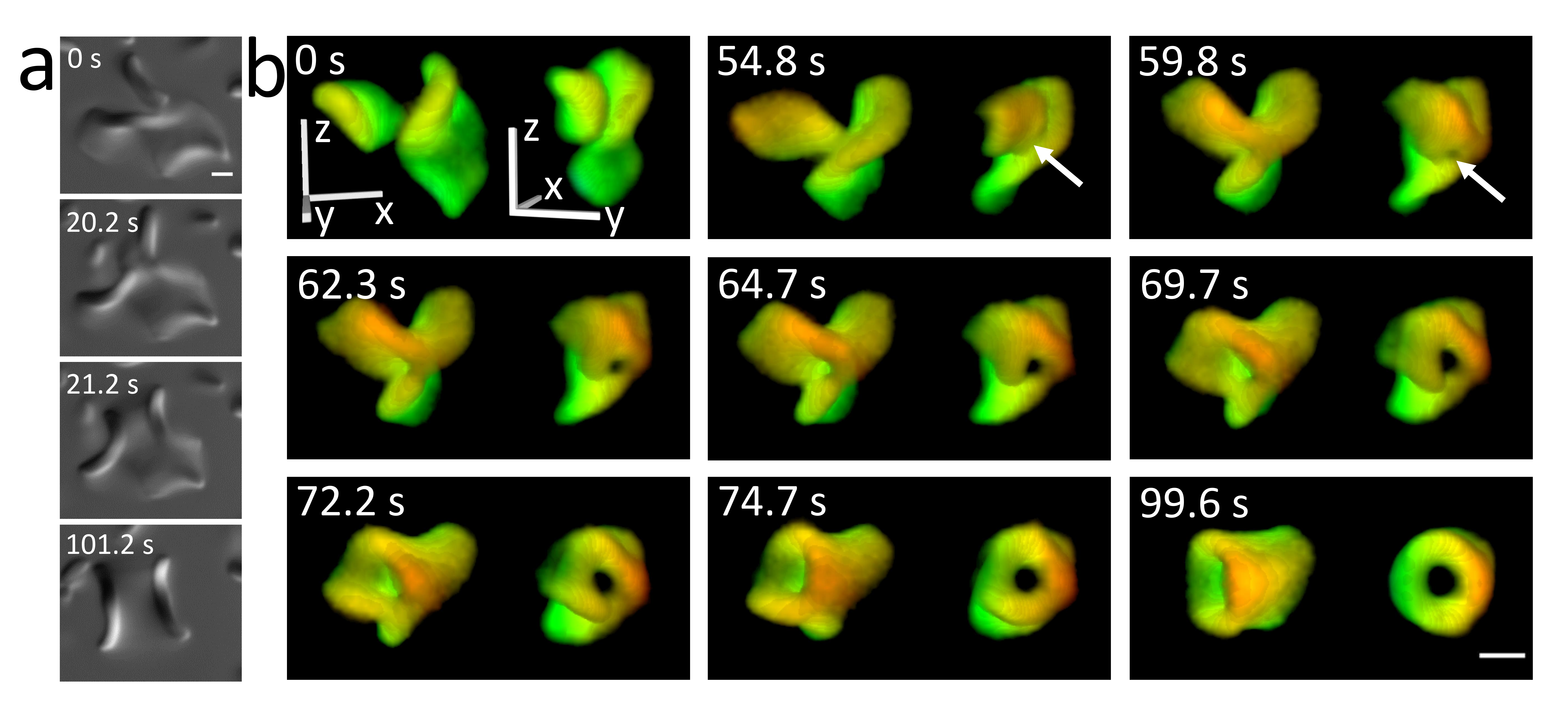}
\caption {\textbf{Saddle-shaped surfaces coalesce to form catenoid-like membranes.} \textbf{(a)} Coalescence of two saddle-shaped membranes, observed with DIC microscopy. \textbf{(b)} Time-lapse 3D false-colored images of the intermediate steps of the coalescence process. These images are obtained by deconvolving $z$-stacks captured using fluorescence microscopy of saddles containing fluorescently-labeled long rods. The event is shown from two orientations, with coordinate axes specified in the first panel. Dextran concentration, $50\,$mg/ml; $n_\textrm{short}=0.2$ and scale bars, 2 \textmu m. The arrow indicates the singularity associated with the formation of a hole.}
\label{fig: saddle_to_cat}
\end{figure*}

Flat disk-like membranes grow in size through lateral coalescence events~\cite{zakhary2014imprintable}. In comparison, saddle-like surfaces exhibit two distinct coalescence pathways. Lateral coalescence events lead to Enneper surfaces of greater area, sometimes of the same order, and sometimes of a higher order. However, we also observed a distinct coalescence pathway that changes the topology of the membrane. A 2D section of two saddle-shaped membranes undergoing coalescence via this pathway is not very revealing (Fig.~\ref{fig: saddle_to_cat}a and Movie~S1). To gain insight, we used the unique features of colloidal membranes that allowed for visualizing the coarsening pathways with unprecedented detail. First, micron-sized colloidal membranes in combination with fast $z$-scanning and 3D deconvolution microscopy enabled visualization of complex 3D surfaces. Second, the large size of colloidal membranes slowed the coalescence dynamics, thus enabling real time visualization. Leveraging these features revealed that the two coalescing saddle-shaped surfaces transition into a topologically distinct catenoid-like shape (Fig.~\ref{fig: saddle_to_cat}b and Movie~S2). 

Multiple observations suggest that the catenoid assembly pathway is conserved. Two saddle surfaces approached each other at an almost right angle and fused at a specific location away from the edge of either membrane. Therefore, the point of initial contact was in the interior of both membranes where the rods are normal to the surface. Following this event, a fusion pore nucleated in the vicinity of the initial contact point. This pore grew to a well-defined size along with the continuous transformation of the object into a catenoid-like shape. In comparison, the edge-to-edge lateral coalescence of two saddle-shaped membranes yielded larger or higher order saddles (Movie~S1,~S3). 

\begin{figure*}[ht!]
\centering
\includegraphics[width=\linewidth]{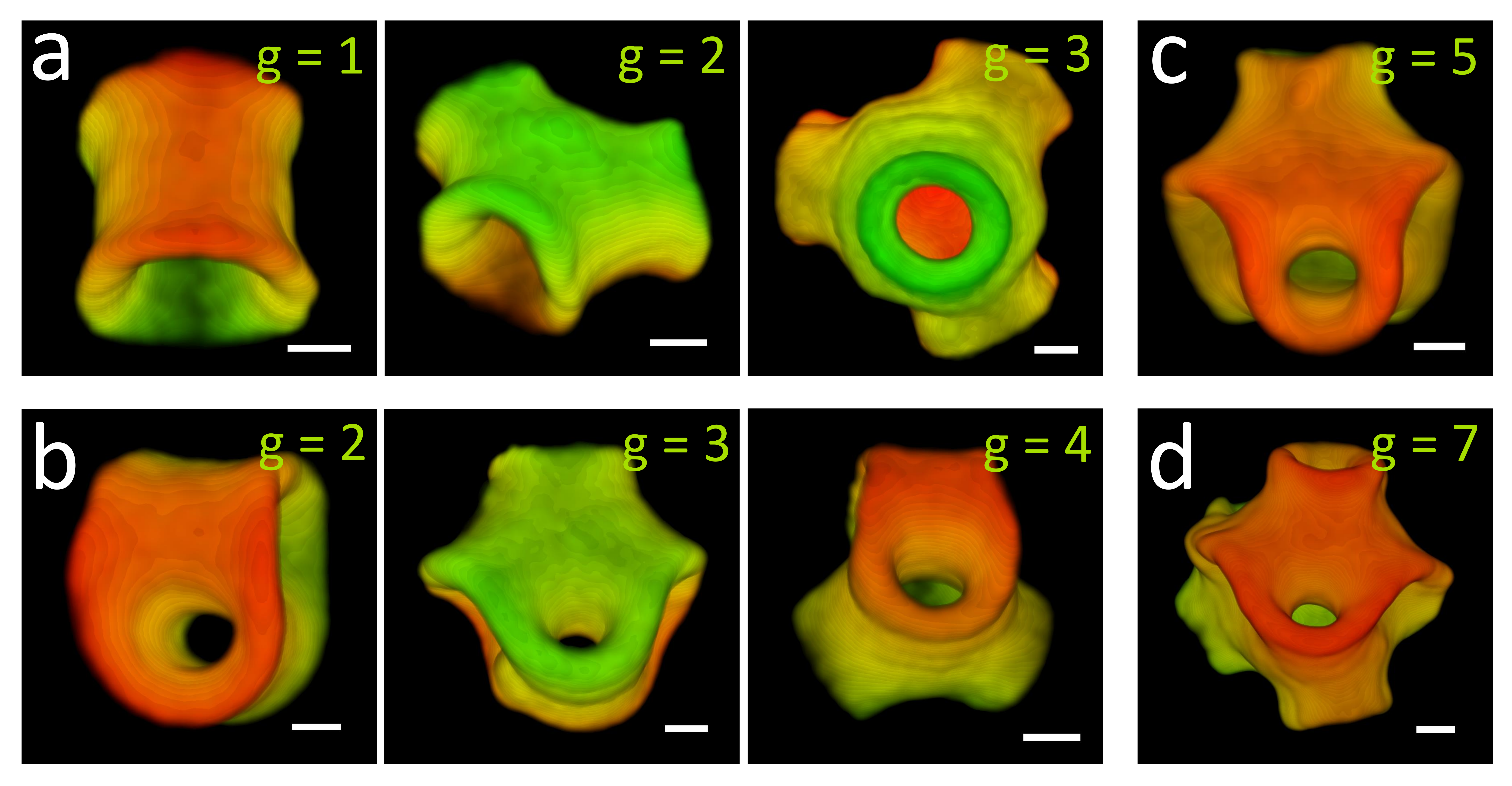}
\caption{\textbf{Topologically complex membranes of different genus.} \textbf{(a)} Catenoid and its derivatives, a 3-noid and a 4-noid. \textbf{(b)} Catenoid with a handle, 3-noid with a handle and 4-noid with a handle. \textbf{(c-d)}  Two complex surfaces of genus 5 and 7. Dextran concentration, 50\,mg/ml; $n_\textrm{short}=0.2$ and scale bars, 2$\,$\textmu m.}
\label{fig: All_genus}
\end{figure*}

\begin{figure*}[ht!]
\centering
\includegraphics[width=0.7\linewidth ]{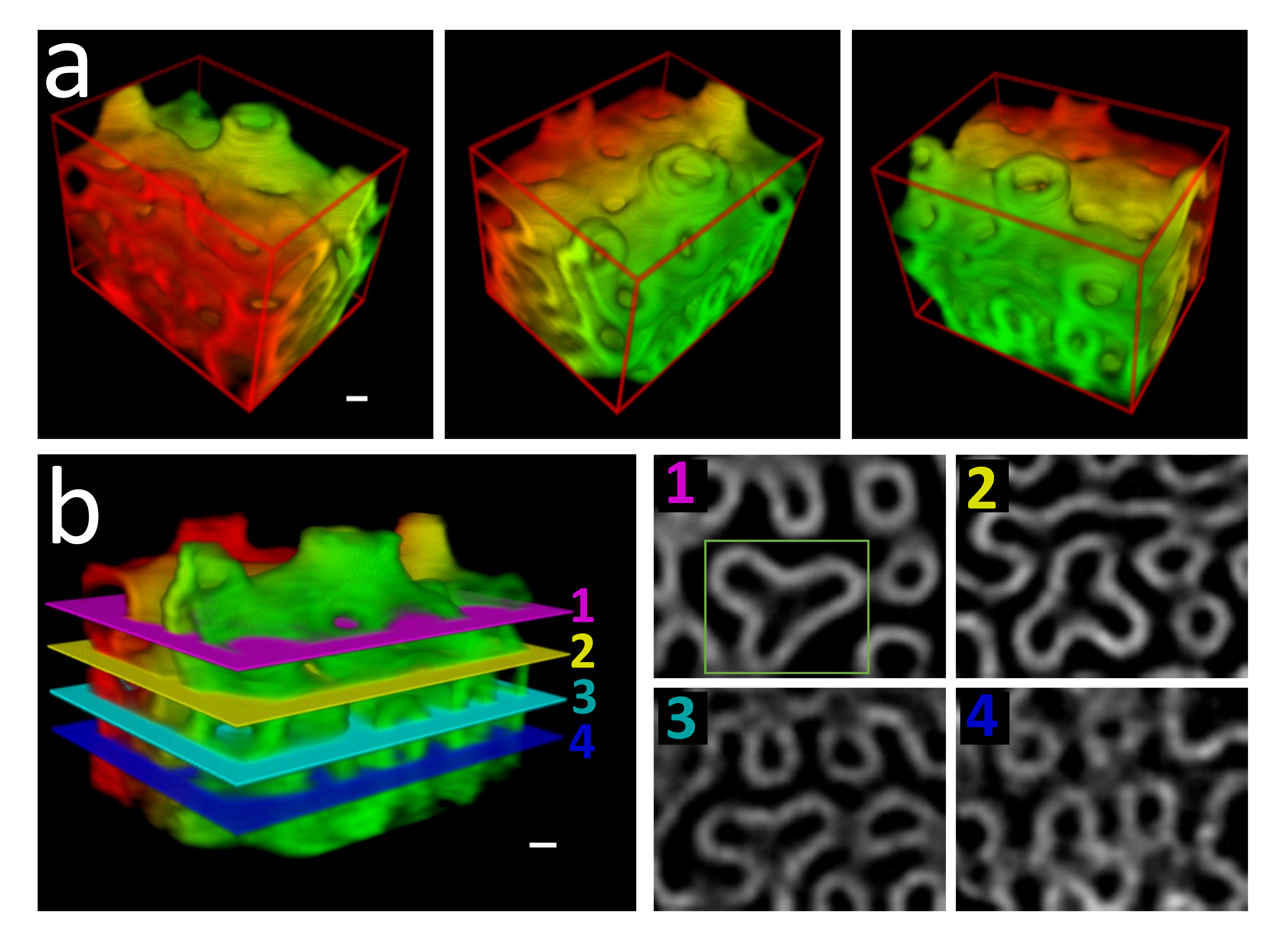}
\caption{\textbf{Colloidal sponge-like phases.} \textbf{(a)} Three different 3D false-colored views of a network-like structure with genus 158. \textbf{(b)} Representative planes within this structure marked as 1, 2, 3 and 4. Grayscale images are 2-D cross-sections of the planes. Rod concentration, 2mg/ml; Dextran concentration, 50 mg/ml and $n_\textrm{short}=0.2$. Scale bars, 2\,\textmu m.}
\label{fig: Network_struc}
\end{figure*}

Catenoid-like membranes were intermediate structures. In the next step, such shapes merged with a saddle-surface to transition into a topologically distinct surface with three openings (Fig.~\ref{fig: All_genus}a, second column). These were similar to 3-noids, which are minimal surfaces having saddle-splay curvature in the form of a catenoid with three openings. The catenoid to 3-noid transition involved nucleation and growth of a fusion pore away from the catenoid edge (Movie~S4). Subsequent coalescence events generated surfaces of increasing complexity (Fig.~\ref{fig: All_genus} and Movie~S5). Besides catenoids and 3-noids we also observed 4-noids (Fig.~\ref{fig: All_genus}a). Well-equilibrated samples showed larger and more complex structures, such as catenoids with many handles (Fig.~\ref{fig: All_genus}b-d). The size of the structures and their topological complexity was limited by the slowing coalescence kinetics. Eventually, Brownian motion of large assemblages slowed down, which reduced further coalescence events. To overcome this limitation, we tilted the sample. Subsequently, intermediate structures slowly sedimented, accumulating at the bottom. There they came in close contact, coalescing further. Such conditions generated macro-scale structures which resembled sponge phases of amphiphilic molecules (Fig.~\ref{fig: Network_struc}a and SI Appendix, Fig.~S4). Confocal microscopy revealed the internal structures of such large-scale assemblages. Regions with quasi-periodicity were found, wherein 3-way open tunnels in each layer rotate by 60° with respect to the consecutive layer (Fig.~\ref{fig: Network_struc}b and Movie~S6). 

We characterized the topology of above-described structures using the genus of the virus-polymer interface, which is equal to the number of doughnut holes in the connected and orientable 3D objects. For example, the saddle structure is topologically equivalent to a sphere (Fig. \ref{fig: saddles_th_exp}a) while the catenoid-like structure is topologically equivalent to a torus (Fig.~\ref{fig: All_genus}a, first column). The 3-noid (Fig.~\ref{fig: All_genus}a, second column) and the handled-catenoid (Fig.~\ref{fig: All_genus}b, first column) are topologically equivalent to a two-torus. The saddle-to-catenoid topological transition changes the genus from 0 to 1. Coalescence of genus 0 and genus 1 structures leads to genus 2 and 3 structures such as 3-noids, 4-noids and other combinations of catenoids with topological handles. The sample spanning structures are more meaningfully characterized by a genus density of 1 hole every 102 \textmu m$^3$, instead of the absolute value of genus.

While genus characterizes the global topology of the self-assembled shapes, spatial maps of mean and Gaussian curvature serve as metrics of local topology. All assemblages had negative Gaussian curvature over large parts of their surface (Fig.~\ref{fig: saddles_th_exp}c and SI Appendix, Fig.~S5). Except for the lowest order saddles, all other complex membranes had finite mean curvature whose spatial variation was connected to the topology  (SI Appendix, Fig.~S5). For example, mean curvature varied along the symmetry axis of a catenoid (SI Appendix, Fig.~S5d). The catenoid with a handle had opposite mean curvature in the two tunnels (SI Appendix, Fig.~S5g). The more complex assemblages such as 3-noid and 4-noid had predominantly negative Gaussian curvature throughout, except for a well-defined region near midsection (SI Appendix, Fig.~S5e-f). 

\begin{figure}[ht!]
\centering
\includegraphics[width=\linewidth ]{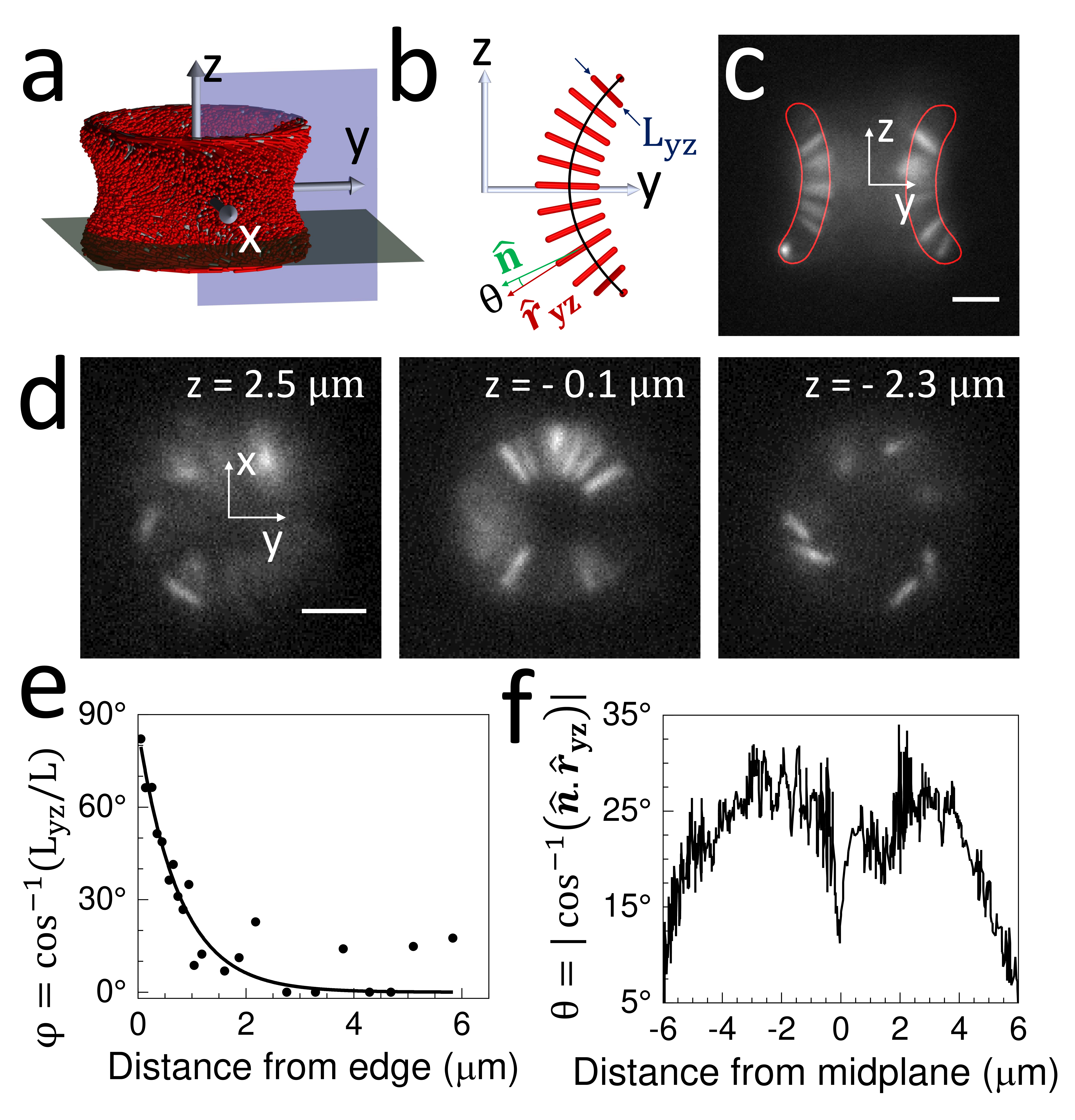}
\caption{\textbf{Orientation of rods in a catenoid.} \textbf{(a)} Schematic of a catenoid showing $x$-$y$ (gray) and $y$-$z$ (violet) planes. \textbf{(b)} Schematic depicting variables that characterize rod orientation. \textbf{(c)} Fluorescence images of individual long rods imaged in $x=0$ plane, with the catenoid wall outline shown in red. \textbf{(d)} Individual long rods imaged at $x$-$y$ planes corresponding to different $z$ values, showing twist in rod orientation. \textbf{(e)} Twist angle $\varphi$, the angle by which the rods tilt away from $y$-$z$ plane, as a function of distance from edge. $L$ is length of the long rods. An exponential fit (black line) to this plot gives a twist penetration depth of 700\,nm. \textbf{(f)} Variation of  $\theta$, the absolute value of angle between $y$-$z$ plane projection of rods, $\hat{r}_{yz}$, and the surface normal of catenoid, $\hat{n}$, with distance from midplane $(z=0)$. $\theta$ is zero for flat membranes, but for catenoids we find a significant deviation from zero.  Dextran concentration is 50\,mg/ml, and $n_\textrm{short}=0.2$ . Scale bars, 2 \textmu m.}
\label{fig: Rod_Ori}
\end{figure}

\subsection*{Microscopic membrane structure} 
Next we investigate how membrane curvature couples to the orientation of the constituent rods with respect to the surface normal. We imaged catenoid-like membranes that were sparsely doped with fluorescently labelled viruses (Fig.~\ref{fig: Rod_Ori}c-d and Movie~S7). Similar to flat membranes~\cite{gibaud2012reconfigurable,barry2008direct}, rods twisted at the exposed catenoid edges. In comparison, close to the catenoid midsection the rods point along the surface normal, generating a radial arrangement (Fig.~\ref{fig: Rod_Ori}d). For our analysis, the catenoid's symmetry axis points along $z$-axis while the circular midsection lies in the $x$-$y$ plane (Fig.~\ref{fig: Rod_Ori}d). We quantified the edge-induced twisting by plotting the angular deviation of the rod axis with respect to the $y$-$z$ plane (Fig.~\ref{fig: Rod_Ori}b), also known as the twist angle. Twist angle decayed exponentially from $\sim\,90^\circ$ to zero with distance from the catenoid edge, with a decay constant of 700\,nm (Fig.~\ref{fig: Rod_Ori}e), comparable to the twist penetration depth in flat membranes~\cite{barry2008direct}. An anomalous behavior is seen when rod orientation in a catenoid is observed in $y$-$z$ midplane; the projection of rods deviates up to 25$^\circ$ from the local surface normal (Fig.~\ref{fig: Rod_Ori}f). This deviation of the rod axis from the local surface normal leads to an additional free energy cost, which might be reduced by the bump which appears in the midsection of large catenoids (SI Appendix, Fig.~S6a). Rod orientation in saddles of order $m=1$ is nearly along the local surface normal everywhere except near the edges (SI Appendix, Fig.~S7).

In principle, short rods could preferentially reside next to the inner or outer membrane surface, or close to the membrane midplane~\cite{siavashpouri2019structure}. By labelling both rods types, we determined the center of mass of both short rod and long rods (SI Appendix, Fig.~S8). Such efforts demonstrate that the center of mass of short rods are preferentially located at the membrane midplane.   

\subsection*{Stimuli induced membrane folding}  
\begin{figure*}[ht!]
\centering
\includegraphics[width=0.85\linewidth ]{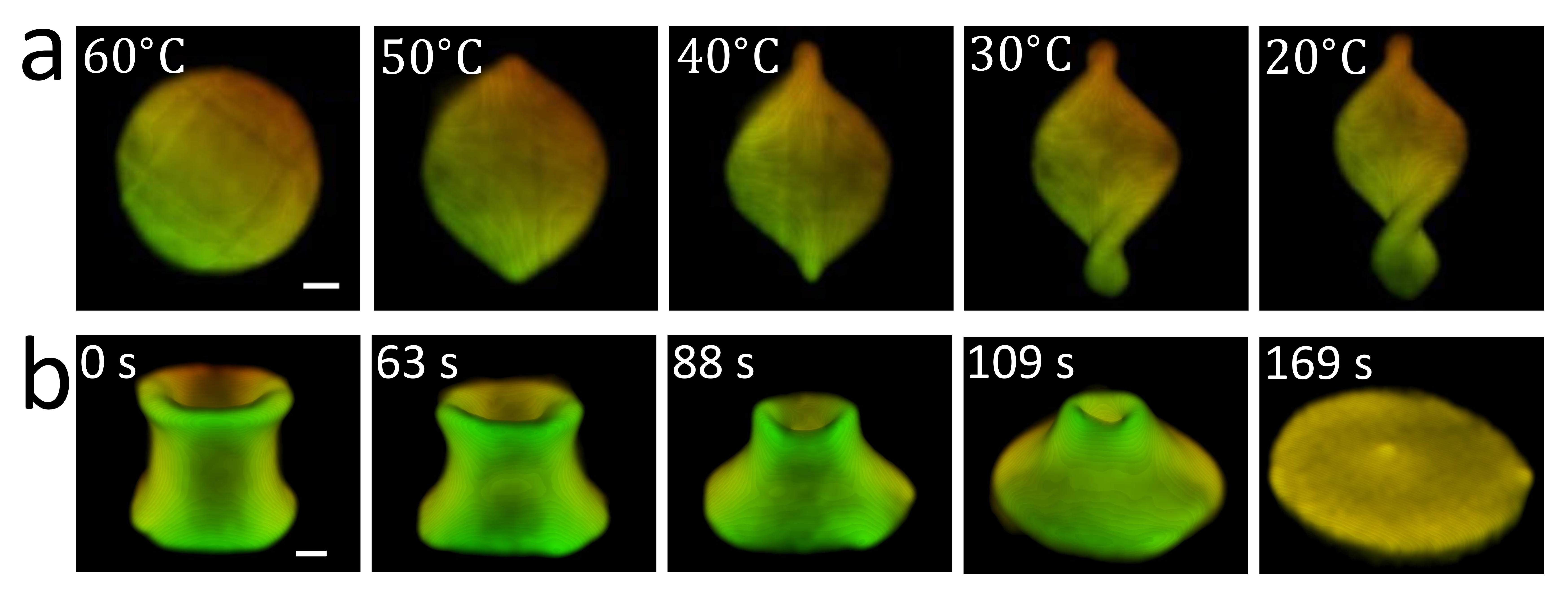}
\caption{ \textbf{Stimuli-responsive colloidal membranes change geometry and topology.} \textbf{(a)} A flat membrane transforms into a twisted saddle or a helicoid upon decreasing temperature, which increases the strength of the depletion attraction. \textbf{(b)} A catenoid transforms into a flat membrane after elevating temperature to $70\,^\circ$C. Time-lapse images show deconvolved z-stacks that were 3D rendered in false-color. The samples were self-assembled at room temperature with $n_\textrm{short}=0.23$ and 15.6\,mg/ml of PEG 35K as depletant dissolved in 175 mM NaCl, 20 mM Tris, pH 8.0 buffer. Scale bars, 2 \textmu m.}
\label{fig: PEG_exp}
\end{figure*}

The phase diagram shows that the disk-to-saddle transition occurs with increasing Dextran concentration or equivalently the osmotic pressure (Fig.~\ref{fig: PhaseDiag_Schematics}d). Instead of Dextran, we assembled membranes with poly(ethylene glycol) (PEG), a polymer whose osmotic pressure exhibits significant temperature dependence~\cite{PEG_osmotic_pressure}. The saddle to disk transition could be induced by changing the sample temperature. We assembled saddle-shaped membranes at room temperature. Elevating the sample temperature to $60\,^\circ$C {\it in situ} decreased the osmotic pressure of the enveloping polymer and concentration of the rods within the membrane. In response, the saddle membrane transformed into a flat disk. The saddle to disk transition was reversible (Movie~S8). On decreasing the temperature, curved regions nucleated near the edge of the flat membrane and eventually a saddle formed (Fig.~\ref{fig: PEG_exp}a and Movie~S9).

Topology influenced the stability of curved surfaces. At $60\,^\circ$C most saddle surfaces transformed into flat disk-shapped membranes. In comparison, topologically distinct catenoids showed almost no shape change, except for a slight increase in the neck radius. Increasing the temperature further to $\sim70\,^\circ$C destabilized the catenoids; they transitioned into disk by a specific kinetic pathway. In the first step, one exposed edge of a catenoid started decreasing in radius, and finally closed. Subsequently, the curved surface transformed into a flat membrane (Fig.~\ref{fig: PEG_exp}b). The catenoid to disk transition was irreversible. 

\subsection*{Estimating Gaussian curvature modulus}
Stability of saddles suggested that the Gaussian modulus of two component membranes is $\bar\kappa \approx 1000$ $k_BT$. A simple model explains this finding. We first note that several independent measurements suggest that single component colloidal membranes have a $\bar\kappa \approx 200$ $k_BT$~\cite{gibaud2017achiral,jia2017chiral,balchunas2020force}. This value can be explained by the increase in the volume available to the depleting polymers when the membrane assumes a saddle-like negative Gaussian curvature, which leads to a positive contribution $\bar{\kappa}_\mathrm{p}$ to the Gaussian curvature modulus, $\bar{\kappa}_\mathrm{p}=(D+d)^3 n k_\mathrm{B}T/12$, where $D$ is the thickness of the colloidal membrane, $d$ is the diameter of the depleting polymer, and $n$ is the number density of the polymers~\cite{gibaud2017achiral}. The membrane itself contributes a negative contribution $\bar{\kappa}_\mathrm{m}$, which is given by the second moment of the in-plane lateral stress $\sigma$ of the flat state, $\bar{\kappa}_\mathrm{m}=\int\sigma z^2\mathrm{d}z$, where $z$ measures distance across the membrane thickness, measured from the midplane~\cite{Harbich_etal1978,Helfrich1981}.

For membranes containing a small number fraction $\alpha$ of short rods of length $D_\mathrm{s}$ among long rods of length $D_\mathrm{l}$, we estimate the second moment of the in-plane stress as
\begin{equation}
    \int\sigma z^2\mathrm{d}z=\alpha\int^{D_\mathrm{s}/2}_{-D_\mathrm{s}/2}\sigma_\mathrm{s}z^2\mathrm{d}z+
    (1-\alpha)\int^{D_\mathrm{l}/2}_{-D_\mathrm{l}/2}\sigma_\mathrm{l}z^2\mathrm{d}z,
\end{equation}
where $\sigma_{\mathrm{s},\mathrm{l}}=-n k_\mathrm{B}T(D_{\mathrm{s},\mathrm{l}}+d)/D_{\mathrm{s},\mathrm{l}}$ is the compressive stress due to the osmotic forces of the polymers that arises in a flat membrane composed of short rods only or long rods only. 

Assuming at small volume fraction $\alpha$ the short rods do not affect the polymeric contribution $\bar{\kappa}_\mathrm{p}$, and that 
$D_{\mathrm{s},\mathrm{l}}\gg d$, we find
\begin{equation}
    \bar{\kappa}=\bar{\kappa}_\mathrm{p}+\bar{\kappa}_\mathrm{m}=\frac{1}{6}nk_\mathrm{B}T D_\mathrm{l}^2d+\frac{\alpha}{12}n k_\mathrm{B}T(D_\mathrm{l}^3-D_\mathrm{s}^3).
\end{equation}
The Gaussian curvature modulus $\bar{\kappa}$ increases linearly with the number fraction of short rods at small number fraction because increasing the number of short rods decreases the negative contribution $\bar{\kappa}_\mathrm{m}$. For Dextran polymers with $d=30$\,nm, a molecular weight of $500,000$\,g/mol,  and a concentration of 50\,mg/mL, we estimate the Gaussian modulus to be $\bar{\kappa}_\mathrm{l}\approx400\,k_\mathrm{B}T$ for a membrane composed of long rods only and $\bar{\kappa}_\mathrm{s}\approx200\,k_\mathrm{B}T$ for a membrane composed of short rods only. At a short rod fraction of $\alpha=0.2$, the correction is $\alpha nk_\mathrm{B}T(D_\mathrm{l}^3-D_\mathrm{s}^3)/12\approx1000\,k_\mathrm{B}T$. We need to mention a caveat in order not to over interpret our simplified model. Specifically, the validity of applying this formula to membranes composed of heterogeneous molecules such as lipids has been called into question~\cite{TerziErgruderDeserno2019}. 

\section*{Discussion and Conclusions}
We described the diverse 3D shapes observed in two-component miscible colloidal membranes. Increasing the fraction of short-rods destabilized flat disk-like membranes, leading to assembly of saddle-shaped Enneper-like surfaces. These intermediate structures continued coalescing with each other to form topologically more complex structures. The micron length scale of colloidal membranes enabled visualization of the multistep coalescence pathway in real time. 

Theoretical analysis suggests that the stability of disk-like membranes is controlled by the Gaussian modulus. Our proposed model is too crude to quantitatively predict the boundaries of the disk-to-saddle transition, but it suggests the following qualitative explanation. The edge tension favors flat disk-like membranes. At low volume fraction of short rods, the intrinsic Gaussian curvature modulus is not high enough to overcome the desire of the edge tension to keep the membrane flat. Increasing the fraction of short rods favored the formation of saddle surfaces. In the saddle regime, the Gaussian modulus is large and positive, thus saddle-shaped surfaces decrease the membrane deformation energy, compensating for the excess edge that is associated with the formation of a non-flat surface. This strongly suggests that increasing the fraction of short rods increases the magnitude of the Gaussian modulus. A simple geometrical argument based on excluded volume explains how fraction of short rods controls the Gaussian curvature modulus. Finally, increasing the fraction of short rods even further induces lateral phase-separation. Thus, the Gaussian curvature modulus drops abruptly, and one recovers flat disk-shaped phase-separated membranes. Previous studies have observed different surfaces with negative Gaussian curvature, namely twisted ribbons~\cite{gibaud2012reconfigurable,balchunas2020force}. Those studies used a virus which had significantly lower edge tension. Extended twisted ribbons have excess edge energy, when compared to the compact structures studied here. 

Colloidal membranes and conventional lipid bilayers are described by the same continuum energy~\cite{helfrich1973elastic}. Thus, besides demonstrating a robust method for shaping colloidal membranes, our results also advance the understanding of all membrane-based materials. Gaussian modulus is a key physical quantity that governs phenomena involving membrane curvature generation and modulation such as endocytosis and exocytosis, cell differentiation, and cell motility \cite{endocytosis1,endocytosis2,exocytosis1,fission1,cellmotility1,cellfunction1}. Conventional lipid bilayer vesicles do not permit easy understanding of how Gaussian modulus affects their morphology, as the Gauss-Bonnet theorem requires that the Gaussian curvature energy integrates to a system independent constant value~\cite{deserno1,kozlov1,kozlov2}. Consequently, measurement and control of Gaussian modulus in lipid bilayers is challenging~\cite{turner1,baumgart1,gauss1}. Open edges of colloidal membranes enabled our study of how  Gaussian modulus influences the stability of flat disk-shaped membranes. Our method of tuning Gaussian modulus by doping membranes with miscible short rods reveal a microscopic mechanism that cells could use for curvature generation and maintenance. 

Ranging from simple crystals and liquid crystals to more exotic polymers, glasses, alloys and diamond-like structures, colloids can form analogs of diverse atomic materials~\cite{colloids1,van1995real,schall2006visualizing,he2020colloidal}. Colloidal length scale provides a unique opportunity to visualize the real space structure and dynamics that is inaccessible in atomic materials. The structures we observed at the end stages of the coalescence are reminiscent of finite-sized cubosomes and gyroid-like phases observed in conventional amphiphilic systems~\cite{Demurtas2015,Cubosome_Barriga,Gazeau_1989,Cates_1988}. The colloidal length scale of our system allowed for detailed real time imaging of the molecular assembly pathways. For example, we have observed how two saddle-shaped surfaces come together to form a topologically distinct catenoid-like structure. Intriguingly, this coalescence proceeds when two surfaces come together at a very specific orientation, and is accompanied by the formation of a  fusion pore  and its subsequent enlargement. A detailed understanding of this conserved dynamical pathway warrants future studies. 

More broadly, many dynamical processes, such as the breakup of fluid droplets, exhibit shape transformations that are characterized by the appearance of  singularities \cite{singularity1,singularity2}. Membrane based processes, such as infection of membrane enveloped viruses, also exhibit singularities as they undergo fission and fusion transformations. However, the molecular length scale of lipid bilayers prevents visualizing the dynamics of these singularities. Our experiments at micron length scales reveal that fusion of saddle-shaped membranes into a catenoid has an intriguing singularity, involving the creation of a fusion pore and its growth.  

Finally, our work provides a new mechanism for topological shaping of thin elastic membrane-like sheets by controlling their Gaussian modulus. This complements the usual technique of folding sheets through mechanical instabilities induced by in-plane differential swelling/shrinkage or application of external confining forces~\cite{klein2007shaping,Huang650,Menon1}. We foresee that spatial control of Gaussian modulus of colloidal membranes would lead to an even richer landscape of topologically complex surfaces.

\section*{Materials and Methods}
M13KO7 and M13-wt virus were grown using host E. coli strain ER2738, following standard biological protocols~\cite{Sambrook}. Gel electrophoresis revealed that the purified M13-wt virus had a significant amount of end-to-end multimers, which prevents defect-free membrane formation. The multimers were removed using isotropic-nematic phase separation~\cite{barry2010entropy}. All viruses were suspended in 100 mM NaCl, 20mM Tris-HCl (pH = 8.0) media.

Viruses were labeled with either DyLight 550 or DyLight 488 (Thermo Fisher Scientific) amine reactive dye for the purpose of fluorescence imaging. There are $\sim$3600 and $\sim$2700 labeling sites on M13KO7 and M13-wt rods, respectively. 10\% of the sites were labeled for experiments that image the motion and orientation of individual rods. 1\% of the sites were labeled for all other experiments. 

The number density of viruses in a given suspension was measured with UV-Vis spectrophotometer (Multiscan GO, Thermo Fisher Scientific). The two kinds of viruses were mixed at the desired stoichiometric ratio, and Dextran (MW 500 kDa; Sigma-Aldrich) was added. Coverslips were coated with polyacrylamide brush before sample preparation to prevent membranes from adhering to coverslips. Sample chambers were made from coated coverslips and cleaned slides, using parafilm as spacer. The suspension was injected into the chamber and the chamber was sealed with optical glue (Norland).

Samples were observed using an inverted widefield microscope (Olympus IX83) equipped with 100X oil immersion phase and DIC objectives (UPLanFLN-100X/1.30 Oil Ph3, UPLanFLN-100X/1.30 Oil), motorized z-drive, CCD and EMCCD cameras (Photometrics CoolSNAP HQ2, Andor iXon Ultra 888). A Peltier stage (PE120, Linkam) was used to vary sample temperature. Z-stacks were captured using this microscope in fluorescence mode, followed by deconvolution to represent the structures in 3D qualitatively. Confocal microscope (Zeiss LSM 880 Airyscan, equipped with Plan Apo 63X 1.4NA oil objective) was used for capturing z-stacks for quantitative analysis.

\begin{acknowledgments}
We thank Megan Kerr for helpful discussions. N.P.M., A.B., and Z.D. acknowledge support by NSF-DMR-1905384. R.A.P. and T.R.P. acknowledge support from National Science Foundation Grant No. NSF CMMI-2020098. We also acknowledge support of NSF MRSEC through grant DMR-2011486 (Z.D., R.A.P., and T.R.P.). A.K. and P.S. acknowledge funding support from SERB, DST Grant No. CRG/2019/000855.  This research was supported in part by the National Science Foundation under Grant No. NSF PHY-1748958. N.P.M. acknowledges support from the Helen Hay Whitney Foundation. 
\end{acknowledgments}

\bibliography{refs}

\begin{thebibliography}{71}%
\makeatletter
\providecommand \@ifxundefined [1]{%
 \@ifx{#1\undefined}
}%
\providecommand \@ifnum [1]{%
 \ifnum #1\expandafter \@firstoftwo
 \else \expandafter \@secondoftwo
 \fi
}%
\providecommand \@ifx [1]{%
 \ifx #1\expandafter \@firstoftwo
 \else \expandafter \@secondoftwo
 \fi
}%
\providecommand \natexlab [1]{#1}%
\providecommand \enquote  [1]{``#1''}%
\providecommand \bibnamefont  [1]{#1}%
\providecommand \bibfnamefont [1]{#1}%
\providecommand \citenamefont [1]{#1}%
\providecommand \href@noop [0]{\@secondoftwo}%
\providecommand \href [0]{\begingroup \@sanitize@url \@href}%
\providecommand \@href[1]{\@@startlink{#1}\@@href}%
\providecommand \@@href[1]{\endgroup#1\@@endlink}%
\providecommand \@sanitize@url [0]{\catcode `\\12\catcode `\$12\catcode
  `\&12\catcode `\#12\catcode `\^12\catcode `\_12\catcode `\%12\relax}%
\providecommand \@@startlink[1]{}%
\providecommand \@@endlink[0]{}%
\providecommand \url  [0]{\begingroup\@sanitize@url \@url }%
\providecommand \@url [1]{\endgroup\@href {#1}{\urlprefix }}%
\providecommand \urlprefix  [0]{URL }%
\providecommand \Eprint [0]{\href }%
\providecommand \doibase [0]{https://doi.org/}%
\providecommand \selectlanguage [0]{\@gobble}%
\providecommand \bibinfo  [0]{\@secondoftwo}%
\providecommand \bibfield  [0]{\@secondoftwo}%
\providecommand \translation [1]{[#1]}%
\providecommand \BibitemOpen [0]{}%
\providecommand \bibitemStop [0]{}%
\providecommand \bibitemNoStop [0]{.\EOS\space}%
\providecommand \EOS [0]{\spacefactor3000\relax}%
\providecommand \BibitemShut  [1]{\csname bibitem#1\endcsname}%
\let\auto@bib@innerbib\@empty
\bibitem [{\citenamefont {Seifert}(1997)}]{ConfMembVesicles}%
  \BibitemOpen
  \bibfield  {author} {\bibinfo {author} {\bibfnamefont {U.}~\bibnamefont
  {Seifert}},\ }\bibfield  {title} {\bibinfo {title} {Configurations of fluid
  membranes and vesicles},\ }\href@noop {} {\bibfield  {journal} {\bibinfo
  {journal} {Advances in Physics}\ }\textbf {\bibinfo {volume} {46}},\ \bibinfo
  {pages} {13} (\bibinfo {year} {1997})}\BibitemShut {NoStop}%
\bibitem [{\citenamefont {Almsherqi}\ \emph {et~al.}(2006)\citenamefont
  {Almsherqi}, \citenamefont {Kohlwein},\ and\ \citenamefont
  {Deng}}]{Almsherqi2006}%
  \BibitemOpen
  \bibfield  {author} {\bibinfo {author} {\bibfnamefont {Z.~A.}\ \bibnamefont
  {Almsherqi}}, \bibinfo {author} {\bibfnamefont {S.~D.}\ \bibnamefont
  {Kohlwein}},\ and\ \bibinfo {author} {\bibfnamefont {Y.}~\bibnamefont
  {Deng}},\ }\bibfield  {title} {\bibinfo {title} {Cubic membranes: a legend
  beyond the {Flatland*} of cell membrane organization},\ }\href@noop {}
  {\bibfield  {journal} {\bibinfo  {journal} {The Journal of cell biology}\
  }\textbf {\bibinfo {volume} {173}},\ \bibinfo {pages} {839} (\bibinfo {year}
  {2006})}\BibitemShut {NoStop}%
\bibitem [{\citenamefont {Michalet}\ and\ \citenamefont
  {Bensimon}(1995)}]{Michalet666}%
  \BibitemOpen
  \bibfield  {author} {\bibinfo {author} {\bibfnamefont {X.}~\bibnamefont
  {Michalet}}\ and\ \bibinfo {author} {\bibfnamefont {D.}~\bibnamefont
  {Bensimon}},\ }\bibfield  {title} {\bibinfo {title} {Observation of stable
  shapes and conformal diffusion in genus 2 vesicles},\ }\href@noop {}
  {\bibfield  {journal} {\bibinfo  {journal} {Science}\ }\textbf {\bibinfo
  {volume} {269}},\ \bibinfo {pages} {666} (\bibinfo {year}
  {1995})}\BibitemShut {NoStop}%
\bibitem [{\citenamefont {Longley}\ and\ \citenamefont
  {McIntosh}(1983)}]{Longley1983}%
  \BibitemOpen
  \bibfield  {author} {\bibinfo {author} {\bibfnamefont {W.}~\bibnamefont
  {Longley}}\ and\ \bibinfo {author} {\bibfnamefont {T.~J.}\ \bibnamefont
  {McIntosh}},\ }\bibfield  {title} {\bibinfo {title} {A bicontinuous
  tetrahedral structure in a liquid-crystalline lipid},\ }\href@noop {}
  {\bibfield  {journal} {\bibinfo  {journal} {Nature}\ }\textbf {\bibinfo
  {volume} {303}},\ \bibinfo {pages} {612} (\bibinfo {year}
  {1983})}\BibitemShut {NoStop}%
\bibitem [{\citenamefont {McMahon}\ and\ \citenamefont
  {Gallop}(2005)}]{McMahon2005}%
  \BibitemOpen
  \bibfield  {author} {\bibinfo {author} {\bibfnamefont {H.~T.}\ \bibnamefont
  {McMahon}}\ and\ \bibinfo {author} {\bibfnamefont {J.~L.}\ \bibnamefont
  {Gallop}},\ }\bibfield  {title} {\bibinfo {title} {Membrane curvature and
  mechanisms of dynamic cell membrane remodelling},\ }\href@noop {} {\bibfield
  {journal} {\bibinfo  {journal} {Nature}\ }\textbf {\bibinfo {volume} {438}},\
  \bibinfo {pages} {590} (\bibinfo {year} {2005})}\BibitemShut {NoStop}%
\bibitem [{\citenamefont {Chernomordik}\ and\ \citenamefont
  {Kozlov}(2008)}]{kozlov1}%
  \BibitemOpen
  \bibfield  {author} {\bibinfo {author} {\bibfnamefont {L.~V.}\ \bibnamefont
  {Chernomordik}}\ and\ \bibinfo {author} {\bibfnamefont {M.~M.}\ \bibnamefont
  {Kozlov}},\ }\bibfield  {title} {\bibinfo {title} {Mechanics of membrane
  fusion},\ }\href@noop {} {\bibfield  {journal} {\bibinfo  {journal} {Nature
  structural \& molecular biology}\ }\textbf {\bibinfo {volume} {15}},\
  \bibinfo {pages} {675} (\bibinfo {year} {2008})}\BibitemShut {NoStop}%
\bibitem [{\citenamefont {Snapp}\ \emph {et~al.}(2003)\citenamefont {Snapp},
  \citenamefont {Hegde}, \citenamefont {Francolini}, \citenamefont {Lombardo},
  \citenamefont {Colombo}, \citenamefont {Pedrazzini}, \citenamefont
  {Borgese},\ and\ \citenamefont {Lippincott-Schwartz}}]{snapp2003formation}%
  \BibitemOpen
  \bibfield  {author} {\bibinfo {author} {\bibfnamefont {E.~L.}\ \bibnamefont
  {Snapp}}, \bibinfo {author} {\bibfnamefont {R.~S.}\ \bibnamefont {Hegde}},
  \bibinfo {author} {\bibfnamefont {M.}~\bibnamefont {Francolini}}, \bibinfo
  {author} {\bibfnamefont {F.}~\bibnamefont {Lombardo}}, \bibinfo {author}
  {\bibfnamefont {S.}~\bibnamefont {Colombo}}, \bibinfo {author} {\bibfnamefont
  {E.}~\bibnamefont {Pedrazzini}}, \bibinfo {author} {\bibfnamefont
  {N.}~\bibnamefont {Borgese}},\ and\ \bibinfo {author} {\bibfnamefont
  {J.}~\bibnamefont {Lippincott-Schwartz}},\ }\bibfield  {title} {\bibinfo
  {title} {Formation of stacked {ER} cisternae by low affinity protein
  interactions},\ }\href@noop {} {\bibfield  {journal} {\bibinfo  {journal}
  {Journal of cell biology}\ }\textbf {\bibinfo {volume} {163}},\ \bibinfo
  {pages} {257} (\bibinfo {year} {2003})}\BibitemShut {NoStop}%
\bibitem [{\citenamefont {Bussi}\ \emph {et~al.}(2019)\citenamefont {Bussi},
  \citenamefont {Shimoni}, \citenamefont {Weiner}, \citenamefont {Kapon},
  \citenamefont {Charuvi}, \citenamefont {Nevo}, \citenamefont {Efrati},\ and\
  \citenamefont {Reich}}]{bussi2019fundamental}%
  \BibitemOpen
  \bibfield  {author} {\bibinfo {author} {\bibfnamefont {Y.}~\bibnamefont
  {Bussi}}, \bibinfo {author} {\bibfnamefont {E.}~\bibnamefont {Shimoni}},
  \bibinfo {author} {\bibfnamefont {A.}~\bibnamefont {Weiner}}, \bibinfo
  {author} {\bibfnamefont {R.}~\bibnamefont {Kapon}}, \bibinfo {author}
  {\bibfnamefont {D.}~\bibnamefont {Charuvi}}, \bibinfo {author} {\bibfnamefont
  {R.}~\bibnamefont {Nevo}}, \bibinfo {author} {\bibfnamefont {E.}~\bibnamefont
  {Efrati}},\ and\ \bibinfo {author} {\bibfnamefont {Z.}~\bibnamefont
  {Reich}},\ }\bibfield  {title} {\bibinfo {title} {Fundamental helical
  geometry consolidates the plant photosynthetic membrane},\ }\href@noop {}
  {\bibfield  {journal} {\bibinfo  {journal} {Proceedings of the National
  Academy of Sciences}\ }\textbf {\bibinfo {volume} {116}},\ \bibinfo {pages}
  {22366} (\bibinfo {year} {2019})}\BibitemShut {NoStop}%
\bibitem [{\citenamefont {Harrison}(2008)}]{harrison2008viral}%
  \BibitemOpen
  \bibfield  {author} {\bibinfo {author} {\bibfnamefont {S.~C.}\ \bibnamefont
  {Harrison}},\ }\bibfield  {title} {\bibinfo {title} {Viral membrane fusion},\
  }\href@noop {} {\bibfield  {journal} {\bibinfo  {journal} {Nature structural
  \& molecular biology}\ }\textbf {\bibinfo {volume} {15}},\ \bibinfo {pages}
  {690} (\bibinfo {year} {2008})}\BibitemShut {NoStop}%
\bibitem [{\citenamefont {Zimmerberg}\ and\ \citenamefont
  {Kozlov}(2006)}]{zimmerberg2006proteins}%
  \BibitemOpen
  \bibfield  {author} {\bibinfo {author} {\bibfnamefont {J.}~\bibnamefont
  {Zimmerberg}}\ and\ \bibinfo {author} {\bibfnamefont {M.~M.}\ \bibnamefont
  {Kozlov}},\ }\bibfield  {title} {\bibinfo {title} {How proteins produce
  cellular membrane curvature},\ }\href@noop {} {\bibfield  {journal} {\bibinfo
   {journal} {Nature reviews Molecular cell biology}\ }\textbf {\bibinfo
  {volume} {7}},\ \bibinfo {pages} {9} (\bibinfo {year} {2006})}\BibitemShut
  {NoStop}%
\bibitem [{\citenamefont {Bykov}\ \emph {et~al.}(2017)\citenamefont {Bykov},
  \citenamefont {Schaffer}, \citenamefont {Dodonova}, \citenamefont {Albert},
  \citenamefont {Plitzko}, \citenamefont {Baumeister}, \citenamefont {Engel},\
  and\ \citenamefont {Briggs}}]{bykov2017structure}%
  \BibitemOpen
  \bibfield  {author} {\bibinfo {author} {\bibfnamefont {Y.~S.}\ \bibnamefont
  {Bykov}}, \bibinfo {author} {\bibfnamefont {M.}~\bibnamefont {Schaffer}},
  \bibinfo {author} {\bibfnamefont {S.~O.}\ \bibnamefont {Dodonova}}, \bibinfo
  {author} {\bibfnamefont {S.}~\bibnamefont {Albert}}, \bibinfo {author}
  {\bibfnamefont {J.~M.}\ \bibnamefont {Plitzko}}, \bibinfo {author}
  {\bibfnamefont {W.}~\bibnamefont {Baumeister}}, \bibinfo {author}
  {\bibfnamefont {B.~D.}\ \bibnamefont {Engel}},\ and\ \bibinfo {author}
  {\bibfnamefont {J.~A.}\ \bibnamefont {Briggs}},\ }\bibfield  {title}
  {\bibinfo {title} {The structure of the {COPI} coat determined within the
  cell},\ }\href@noop {} {\bibfield  {journal} {\bibinfo  {journal} {Elife}\
  }\textbf {\bibinfo {volume} {6}},\ \bibinfo {pages} {e32493} (\bibinfo {year}
  {2017})}\BibitemShut {NoStop}%
\bibitem [{\citenamefont {Lewicka}\ and\ \citenamefont
  {Mahadevan}(2021)}]{lewicka2021geometry}%
  \BibitemOpen
  \bibfield  {author} {\bibinfo {author} {\bibfnamefont {M.}~\bibnamefont
  {Lewicka}}\ and\ \bibinfo {author} {\bibfnamefont {L.}~\bibnamefont
  {Mahadevan}},\ }\bibfield  {title} {\bibinfo {title} {Geometry, analysis and
  morphogenesis: {P}roblems and prospects},\ }\href@noop {} {\bibfield
  {journal} {\bibinfo  {journal} {arXiv preprint arXiv:2104.08988}\ } (\bibinfo
  {year} {2021})}\BibitemShut {NoStop}%
\bibitem [{\citenamefont {Savin}\ \emph {et~al.}(2011)\citenamefont {Savin},
  \citenamefont {Kurpios}, \citenamefont {Shyer}, \citenamefont {Florescu},
  \citenamefont {Liang}, \citenamefont {Mahadevan},\ and\ \citenamefont
  {Tabin}}]{savin2011growth}%
  \BibitemOpen
  \bibfield  {author} {\bibinfo {author} {\bibfnamefont {T.}~\bibnamefont
  {Savin}}, \bibinfo {author} {\bibfnamefont {N.~A.}\ \bibnamefont {Kurpios}},
  \bibinfo {author} {\bibfnamefont {A.~E.}\ \bibnamefont {Shyer}}, \bibinfo
  {author} {\bibfnamefont {P.}~\bibnamefont {Florescu}}, \bibinfo {author}
  {\bibfnamefont {H.}~\bibnamefont {Liang}}, \bibinfo {author} {\bibfnamefont
  {L.}~\bibnamefont {Mahadevan}},\ and\ \bibinfo {author} {\bibfnamefont
  {C.~J.}\ \bibnamefont {Tabin}},\ }\bibfield  {title} {\bibinfo {title} {On
  the growth and form of the gut},\ }\href@noop {} {\bibfield  {journal}
  {\bibinfo  {journal} {Nature}\ }\textbf {\bibinfo {volume} {476}},\ \bibinfo
  {pages} {57} (\bibinfo {year} {2011})}\BibitemShut {NoStop}%
\bibitem [{\citenamefont {van Rees}\ \emph {et~al.}(2017)\citenamefont {van
  Rees}, \citenamefont {Vouga},\ and\ \citenamefont
  {Mahadevan}}]{van2017growth}%
  \BibitemOpen
  \bibfield  {author} {\bibinfo {author} {\bibfnamefont {W.~M.}\ \bibnamefont
  {van Rees}}, \bibinfo {author} {\bibfnamefont {E.}~\bibnamefont {Vouga}},\
  and\ \bibinfo {author} {\bibfnamefont {L.}~\bibnamefont {Mahadevan}},\
  }\bibfield  {title} {\bibinfo {title} {Growth patterns for shape-shifting
  elastic bilayers},\ }\href@noop {} {\bibfield  {journal} {\bibinfo  {journal}
  {Proceedings of the National Academy of Sciences}\ }\textbf {\bibinfo
  {volume} {114}},\ \bibinfo {pages} {11597} (\bibinfo {year}
  {2017})}\BibitemShut {NoStop}%
\bibitem [{\citenamefont {Mitchell}\ \emph {et~al.}(2021)\citenamefont
  {Mitchell}, \citenamefont {Cislo}, \citenamefont {Shankar}, \citenamefont
  {Lin}, \citenamefont {Shraiman},\ and\ \citenamefont
  {Streichan}}]{mitchell2021visceral}%
  \BibitemOpen
  \bibfield  {author} {\bibinfo {author} {\bibfnamefont {N.~P.}\ \bibnamefont
  {Mitchell}}, \bibinfo {author} {\bibfnamefont {D.}~\bibnamefont {Cislo}},
  \bibinfo {author} {\bibfnamefont {S.}~\bibnamefont {Shankar}}, \bibinfo
  {author} {\bibfnamefont {Y.}~\bibnamefont {Lin}}, \bibinfo {author}
  {\bibfnamefont {B.~I.}\ \bibnamefont {Shraiman}},\ and\ \bibinfo {author}
  {\bibfnamefont {S.~J.}\ \bibnamefont {Streichan}},\ }\bibfield  {title}
  {\bibinfo {title} {Visceral organ morphogenesis via calcium-patterned muscle
  contractions},\ }\href@noop {} {\bibfield  {journal} {\bibinfo  {journal}
  {bioRxiv}\ } (\bibinfo {year} {2021})}\BibitemShut {NoStop}%
\bibitem [{\citenamefont {Karzbrun}\ \emph {et~al.}(2021)\citenamefont
  {Karzbrun}, \citenamefont {Khankhel}, \citenamefont {Megale}, \citenamefont
  {Glasauer}, \citenamefont {Wyle}, \citenamefont {Britton}, \citenamefont
  {Warmflash}, \citenamefont {Kosik}, \citenamefont {Siggia}, \citenamefont
  {Shraiman} \emph {et~al.}}]{karzbrun2021human}%
  \BibitemOpen
  \bibfield  {author} {\bibinfo {author} {\bibfnamefont {E.}~\bibnamefont
  {Karzbrun}}, \bibinfo {author} {\bibfnamefont {A.~H.}\ \bibnamefont
  {Khankhel}}, \bibinfo {author} {\bibfnamefont {H.~C.}\ \bibnamefont
  {Megale}}, \bibinfo {author} {\bibfnamefont {S.~M.}\ \bibnamefont
  {Glasauer}}, \bibinfo {author} {\bibfnamefont {Y.}~\bibnamefont {Wyle}},
  \bibinfo {author} {\bibfnamefont {G.}~\bibnamefont {Britton}}, \bibinfo
  {author} {\bibfnamefont {A.}~\bibnamefont {Warmflash}}, \bibinfo {author}
  {\bibfnamefont {K.~S.}\ \bibnamefont {Kosik}}, \bibinfo {author}
  {\bibfnamefont {E.~D.}\ \bibnamefont {Siggia}}, \bibinfo {author}
  {\bibfnamefont {B.~I.}\ \bibnamefont {Shraiman}}, \emph {et~al.},\ }\bibfield
   {title} {\bibinfo {title} {Human neural tube morphogenesis in vitro by
  geometric constraints},\ }\href@noop {} {\bibfield  {journal} {\bibinfo
  {journal} {Nature}\ }\textbf {\bibinfo {volume} {599}},\ \bibinfo {pages}
  {268} (\bibinfo {year} {2021})}\BibitemShut {NoStop}%
\bibitem [{\citenamefont {Metzger}\ \emph {et~al.}(2008)\citenamefont
  {Metzger}, \citenamefont {Klein}, \citenamefont {Martin},\ and\ \citenamefont
  {Krasnow}}]{metzger2008branching}%
  \BibitemOpen
  \bibfield  {author} {\bibinfo {author} {\bibfnamefont {R.~J.}\ \bibnamefont
  {Metzger}}, \bibinfo {author} {\bibfnamefont {O.~D.}\ \bibnamefont {Klein}},
  \bibinfo {author} {\bibfnamefont {G.~R.}\ \bibnamefont {Martin}},\ and\
  \bibinfo {author} {\bibfnamefont {M.~A.}\ \bibnamefont {Krasnow}},\
  }\bibfield  {title} {\bibinfo {title} {The branching programme of mouse lung
  development},\ }\href@noop {} {\bibfield  {journal} {\bibinfo  {journal}
  {Nature}\ }\textbf {\bibinfo {volume} {453}},\ \bibinfo {pages} {745}
  (\bibinfo {year} {2008})}\BibitemShut {NoStop}%
\bibitem [{\citenamefont {Cerda}\ and\ \citenamefont
  {Mahadevan}(2003)}]{PhysRevLett.90.074302}%
  \BibitemOpen
  \bibfield  {author} {\bibinfo {author} {\bibfnamefont {E.}~\bibnamefont
  {Cerda}}\ and\ \bibinfo {author} {\bibfnamefont {L.}~\bibnamefont
  {Mahadevan}},\ }\bibfield  {title} {\bibinfo {title} {Geometry and physics of
  wrinkling},\ }\href@noop {} {\bibfield  {journal} {\bibinfo  {journal} {Phys.
  Rev. Lett.}\ }\textbf {\bibinfo {volume} {90}},\ \bibinfo {pages} {074302}
  (\bibinfo {year} {2003})}\BibitemShut {NoStop}%
\bibitem [{\citenamefont {Kim}\ \emph {et~al.}(2012)\citenamefont {Kim},
  \citenamefont {Hanna}, \citenamefont {Byun}, \citenamefont {Santangelo},\
  and\ \citenamefont {Hayward}}]{Kim1201}%
  \BibitemOpen
  \bibfield  {author} {\bibinfo {author} {\bibfnamefont {J.}~\bibnamefont
  {Kim}}, \bibinfo {author} {\bibfnamefont {J.~A.}\ \bibnamefont {Hanna}},
  \bibinfo {author} {\bibfnamefont {M.}~\bibnamefont {Byun}}, \bibinfo {author}
  {\bibfnamefont {C.~D.}\ \bibnamefont {Santangelo}},\ and\ \bibinfo {author}
  {\bibfnamefont {R.~C.}\ \bibnamefont {Hayward}},\ }\bibfield  {title}
  {\bibinfo {title} {Designing responsive buckled surfaces by halftone gel
  lithography},\ }\href@noop {} {\bibfield  {journal} {\bibinfo  {journal}
  {Science}\ }\textbf {\bibinfo {volume} {335}},\ \bibinfo {pages} {1201}
  (\bibinfo {year} {2012})}\BibitemShut {NoStop}%
\bibitem [{\citenamefont {Klein}\ \emph {et~al.}(2007)\citenamefont {Klein},
  \citenamefont {Efrati},\ and\ \citenamefont {Sharon}}]{klein2007shaping}%
  \BibitemOpen
  \bibfield  {author} {\bibinfo {author} {\bibfnamefont {Y.}~\bibnamefont
  {Klein}}, \bibinfo {author} {\bibfnamefont {E.}~\bibnamefont {Efrati}},\ and\
  \bibinfo {author} {\bibfnamefont {E.}~\bibnamefont {Sharon}},\ }\bibfield
  {title} {\bibinfo {title} {Shaping of elastic sheets by prescription of
  non-{E}uclidean metrics},\ }\href@noop {} {\bibfield  {journal} {\bibinfo
  {journal} {Science}\ }\textbf {\bibinfo {volume} {315}},\ \bibinfo {pages}
  {1116} (\bibinfo {year} {2007})}\BibitemShut {NoStop}%
\bibitem [{\citenamefont {Huang}\ \emph {et~al.}(2007)\citenamefont {Huang},
  \citenamefont {Juszkiewicz}, \citenamefont {de~Jeu}, \citenamefont {Cerda},
  \citenamefont {Emrick}, \citenamefont {Menon},\ and\ \citenamefont
  {Russell}}]{Huang650}%
  \BibitemOpen
  \bibfield  {author} {\bibinfo {author} {\bibfnamefont {J.}~\bibnamefont
  {Huang}}, \bibinfo {author} {\bibfnamefont {M.}~\bibnamefont {Juszkiewicz}},
  \bibinfo {author} {\bibfnamefont {W.~H.}\ \bibnamefont {de~Jeu}}, \bibinfo
  {author} {\bibfnamefont {E.}~\bibnamefont {Cerda}}, \bibinfo {author}
  {\bibfnamefont {T.}~\bibnamefont {Emrick}}, \bibinfo {author} {\bibfnamefont
  {N.}~\bibnamefont {Menon}},\ and\ \bibinfo {author} {\bibfnamefont {T.~P.}\
  \bibnamefont {Russell}},\ }\bibfield  {title} {\bibinfo {title} {Capillary
  wrinkling of floating thin polymer films},\ }\href@noop {} {\bibfield
  {journal} {\bibinfo  {journal} {Science}\ }\textbf {\bibinfo {volume}
  {317}},\ \bibinfo {pages} {650} (\bibinfo {year} {2007})}\BibitemShut
  {NoStop}%
\bibitem [{\citenamefont {Barry}\ and\ \citenamefont
  {Dogic}(2010)}]{barry2010entropy}%
  \BibitemOpen
  \bibfield  {author} {\bibinfo {author} {\bibfnamefont {E.}~\bibnamefont
  {Barry}}\ and\ \bibinfo {author} {\bibfnamefont {Z.}~\bibnamefont {Dogic}},\
  }\bibfield  {title} {\bibinfo {title} {Entropy driven self-assembly of
  nonamphiphilic colloidal membranes},\ }\href@noop {} {\bibfield  {journal}
  {\bibinfo  {journal} {Proceedings of the National Academy of Sciences}\
  }\textbf {\bibinfo {volume} {107}},\ \bibinfo {pages} {10348} (\bibinfo
  {year} {2010})}\BibitemShut {NoStop}%
\bibitem [{\citenamefont {Yang}\ \emph {et~al.}(2012)\citenamefont {Yang},
  \citenamefont {Barry}, \citenamefont {Dogic},\ and\ \citenamefont
  {Hagan}}]{yang2012self}%
  \BibitemOpen
  \bibfield  {author} {\bibinfo {author} {\bibfnamefont {Y.}~\bibnamefont
  {Yang}}, \bibinfo {author} {\bibfnamefont {E.}~\bibnamefont {Barry}},
  \bibinfo {author} {\bibfnamefont {Z.}~\bibnamefont {Dogic}},\ and\ \bibinfo
  {author} {\bibfnamefont {M.~F.}\ \bibnamefont {Hagan}},\ }\bibfield  {title}
  {\bibinfo {title} {Self-assembly of {2D} membranes from mixtures of hard rods
  and depleting polymers},\ }\href@noop {} {\bibfield  {journal} {\bibinfo
  {journal} {Soft Matter}\ }\textbf {\bibinfo {volume} {8}},\ \bibinfo {pages}
  {707} (\bibinfo {year} {2012})}\BibitemShut {NoStop}%
\bibitem [{\citenamefont {Balchunas}\ \emph
  {et~al.}(2019{\natexlab{a}})\citenamefont {Balchunas}, \citenamefont
  {Cabanas}, \citenamefont {Zakhary}, \citenamefont {Gibaud}, \citenamefont
  {Fraden}, \citenamefont {Sharma}, \citenamefont {Hagan},\ and\ \citenamefont
  {Dogic}}]{Balchunas}%
  \BibitemOpen
  \bibfield  {author} {\bibinfo {author} {\bibfnamefont {A.~J.}\ \bibnamefont
  {Balchunas}}, \bibinfo {author} {\bibfnamefont {R.~A.}\ \bibnamefont
  {Cabanas}}, \bibinfo {author} {\bibfnamefont {M.~J.}\ \bibnamefont
  {Zakhary}}, \bibinfo {author} {\bibfnamefont {T.}~\bibnamefont {Gibaud}},
  \bibinfo {author} {\bibfnamefont {S.}~\bibnamefont {Fraden}}, \bibinfo
  {author} {\bibfnamefont {P.}~\bibnamefont {Sharma}}, \bibinfo {author}
  {\bibfnamefont {M.~F.}\ \bibnamefont {Hagan}},\ and\ \bibinfo {author}
  {\bibfnamefont {Z.}~\bibnamefont {Dogic}},\ }\bibfield  {title} {\bibinfo
  {title} {Equation of state of colloidal membranes},\ }\href@noop {}
  {\bibfield  {journal} {\bibinfo  {journal} {Soft Matter}\ }\textbf {\bibinfo
  {volume} {15}},\ \bibinfo {pages} {6791} (\bibinfo {year}
  {2019}{\natexlab{a}})}\BibitemShut {NoStop}%
\bibitem [{\citenamefont {Asakura}\ and\ \citenamefont
  {Oosawa}(1954)}]{Asakura_Interaction_2_bodies}%
  \BibitemOpen
  \bibfield  {author} {\bibinfo {author} {\bibfnamefont {S.}~\bibnamefont
  {Asakura}}\ and\ \bibinfo {author} {\bibfnamefont {F.}~\bibnamefont
  {Oosawa}},\ }\bibfield  {title} {\bibinfo {title} {On interaction between two
  bodies immersed in a solution of macromolecules},\ }\href@noop {} {\bibfield
  {journal} {\bibinfo  {journal} {The Journal of Chemical Physics}\ }\textbf
  {\bibinfo {volume} {22}},\ \bibinfo {pages} {1255} (\bibinfo {year}
  {1954})}\BibitemShut {NoStop}%
\bibitem [{\citenamefont {Helfrich}(1973)}]{helfrich1973elastic}%
  \BibitemOpen
  \bibfield  {author} {\bibinfo {author} {\bibfnamefont {W.}~\bibnamefont
  {Helfrich}},\ }\bibfield  {title} {\bibinfo {title} {Elastic properties of
  lipid bilayers: {T}heory and possible experiments},\ }\href@noop {}
  {\bibfield  {journal} {\bibinfo  {journal} {Zeitschrift f{\"u}r
  Naturforschung C}\ }\textbf {\bibinfo {volume} {28}},\ \bibinfo {pages} {693}
  (\bibinfo {year} {1973})}\BibitemShut {NoStop}%
\bibitem [{\citenamefont {Gibaud}\ \emph {et~al.}(2017)\citenamefont {Gibaud},
  \citenamefont {Kaplan}, \citenamefont {Sharma}, \citenamefont {Zakhary},
  \citenamefont {Ward}, \citenamefont {Oldenbourg}, \citenamefont {Meyer},
  \citenamefont {Kamien}, \citenamefont {Powers},\ and\ \citenamefont
  {Dogic}}]{gibaud2017achiral}%
  \BibitemOpen
  \bibfield  {author} {\bibinfo {author} {\bibfnamefont {T.}~\bibnamefont
  {Gibaud}}, \bibinfo {author} {\bibfnamefont {C.~N.}\ \bibnamefont {Kaplan}},
  \bibinfo {author} {\bibfnamefont {P.}~\bibnamefont {Sharma}}, \bibinfo
  {author} {\bibfnamefont {M.~J.}\ \bibnamefont {Zakhary}}, \bibinfo {author}
  {\bibfnamefont {A.}~\bibnamefont {Ward}}, \bibinfo {author} {\bibfnamefont
  {R.}~\bibnamefont {Oldenbourg}}, \bibinfo {author} {\bibfnamefont {R.~B.}\
  \bibnamefont {Meyer}}, \bibinfo {author} {\bibfnamefont {R.~D.}\ \bibnamefont
  {Kamien}}, \bibinfo {author} {\bibfnamefont {T.~R.}\ \bibnamefont {Powers}},\
  and\ \bibinfo {author} {\bibfnamefont {Z.}~\bibnamefont {Dogic}},\ }\bibfield
   {title} {\bibinfo {title} {Achiral symmetry breaking and positive {G}aussian
  modulus lead to scalloped colloidal membranes},\ }\href@noop {} {\bibfield
  {journal} {\bibinfo  {journal} {Proceedings of the National Academy of
  Sciences}\ }\textbf {\bibinfo {volume} {114}},\ \bibinfo {pages} {E3376}
  (\bibinfo {year} {2017})}\BibitemShut {NoStop}%
\bibitem [{\citenamefont {Jia}\ \emph {et~al.}(2017)\citenamefont {Jia},
  \citenamefont {Zakhary}, \citenamefont {Dogic}, \citenamefont {Pelcovits},\
  and\ \citenamefont {Powers}}]{jia2017chiral}%
  \BibitemOpen
  \bibfield  {author} {\bibinfo {author} {\bibfnamefont {L.~L.}\ \bibnamefont
  {Jia}}, \bibinfo {author} {\bibfnamefont {M.~J.}\ \bibnamefont {Zakhary}},
  \bibinfo {author} {\bibfnamefont {Z.}~\bibnamefont {Dogic}}, \bibinfo
  {author} {\bibfnamefont {R.~A.}\ \bibnamefont {Pelcovits}},\ and\ \bibinfo
  {author} {\bibfnamefont {T.~R.}\ \bibnamefont {Powers}},\ }\bibfield  {title}
  {\bibinfo {title} {Chiral edge fluctuations of colloidal membranes},\
  }\href@noop {} {\bibfield  {journal} {\bibinfo  {journal} {Physical Review
  E}\ }\textbf {\bibinfo {volume} {95}},\ \bibinfo {pages} {060701} (\bibinfo
  {year} {2017})}\BibitemShut {NoStop}%
\bibitem [{\citenamefont {Gibaud}\ \emph {et~al.}(2012)\citenamefont {Gibaud},
  \citenamefont {Barry}, \citenamefont {Zakhary}, \citenamefont {Henglin},
  \citenamefont {Ward}, \citenamefont {Yang}, \citenamefont {Berciu},
  \citenamefont {Oldenbourg}, \citenamefont {Hagan}, \citenamefont {Nicastro}
  \emph {et~al.}}]{gibaud2012reconfigurable}%
  \BibitemOpen
  \bibfield  {author} {\bibinfo {author} {\bibfnamefont {T.}~\bibnamefont
  {Gibaud}}, \bibinfo {author} {\bibfnamefont {E.}~\bibnamefont {Barry}},
  \bibinfo {author} {\bibfnamefont {M.~J.}\ \bibnamefont {Zakhary}}, \bibinfo
  {author} {\bibfnamefont {M.}~\bibnamefont {Henglin}}, \bibinfo {author}
  {\bibfnamefont {A.}~\bibnamefont {Ward}}, \bibinfo {author} {\bibfnamefont
  {Y.}~\bibnamefont {Yang}}, \bibinfo {author} {\bibfnamefont {C.}~\bibnamefont
  {Berciu}}, \bibinfo {author} {\bibfnamefont {R.}~\bibnamefont {Oldenbourg}},
  \bibinfo {author} {\bibfnamefont {M.~F.}\ \bibnamefont {Hagan}}, \bibinfo
  {author} {\bibfnamefont {D.}~\bibnamefont {Nicastro}}, \emph {et~al.},\
  }\bibfield  {title} {\bibinfo {title} {Reconfigurable self-assembly through
  chiral control of interfacial tension},\ }\href@noop {} {\bibfield  {journal}
  {\bibinfo  {journal} {Nature}\ }\textbf {\bibinfo {volume} {481}},\ \bibinfo
  {pages} {348} (\bibinfo {year} {2012})}\BibitemShut {NoStop}%
\bibitem [{\citenamefont {Barry}\ \emph {et~al.}(2008)\citenamefont {Barry},
  \citenamefont {Dogic}, \citenamefont {Meyer}, \citenamefont {Pelcovits},\
  and\ \citenamefont {Oldenbourg}}]{barry2008direct}%
  \BibitemOpen
  \bibfield  {author} {\bibinfo {author} {\bibfnamefont {E.}~\bibnamefont
  {Barry}}, \bibinfo {author} {\bibfnamefont {Z.}~\bibnamefont {Dogic}},
  \bibinfo {author} {\bibfnamefont {R.~B.}\ \bibnamefont {Meyer}}, \bibinfo
  {author} {\bibfnamefont {R.~A.}\ \bibnamefont {Pelcovits}},\ and\ \bibinfo
  {author} {\bibfnamefont {R.}~\bibnamefont {Oldenbourg}},\ }\bibfield  {title}
  {\bibinfo {title} {Direct measurement of the twist penetration length in a
  single smectic-{A} layer of colloidal virus particles},\ }\href@noop {}
  {\bibfield  {journal} {\bibinfo  {journal} {The Journal of Physical Chemistry
  B}\ }\textbf {\bibinfo {volume} {113}},\ \bibinfo {pages} {3910} (\bibinfo
  {year} {2008})}\BibitemShut {NoStop}%
\bibitem [{\citenamefont {Sharma}\ \emph {et~al.}(2014)\citenamefont {Sharma},
  \citenamefont {Ward}, \citenamefont {Gibaud}, \citenamefont {Hagan},\ and\
  \citenamefont {Dogic}}]{sharma2014hierarchical}%
  \BibitemOpen
  \bibfield  {author} {\bibinfo {author} {\bibfnamefont {P.}~\bibnamefont
  {Sharma}}, \bibinfo {author} {\bibfnamefont {A.}~\bibnamefont {Ward}},
  \bibinfo {author} {\bibfnamefont {T.}~\bibnamefont {Gibaud}}, \bibinfo
  {author} {\bibfnamefont {M.~F.}\ \bibnamefont {Hagan}},\ and\ \bibinfo
  {author} {\bibfnamefont {Z.}~\bibnamefont {Dogic}},\ }\bibfield  {title}
  {\bibinfo {title} {Hierarchical organization of chiral rafts in colloidal
  membranes},\ }\href@noop {} {\bibfield  {journal} {\bibinfo  {journal}
  {Nature}\ }\textbf {\bibinfo {volume} {513}},\ \bibinfo {pages} {77}
  (\bibinfo {year} {2014})}\BibitemShut {NoStop}%
\bibitem [{\citenamefont {Miller}\ \emph {et~al.}(2019)\citenamefont {Miller},
  \citenamefont {Joshi}, \citenamefont {Sharma}, \citenamefont {Baskaran},
  \citenamefont {Baskaran}, \citenamefont {Grason}, \citenamefont {Hagan},\
  and\ \citenamefont {Dogic}}]{miller2019conformational}%
  \BibitemOpen
  \bibfield  {author} {\bibinfo {author} {\bibfnamefont {J.~M.}\ \bibnamefont
  {Miller}}, \bibinfo {author} {\bibfnamefont {C.}~\bibnamefont {Joshi}},
  \bibinfo {author} {\bibfnamefont {P.}~\bibnamefont {Sharma}}, \bibinfo
  {author} {\bibfnamefont {A.}~\bibnamefont {Baskaran}}, \bibinfo {author}
  {\bibfnamefont {A.}~\bibnamefont {Baskaran}}, \bibinfo {author}
  {\bibfnamefont {G.~M.}\ \bibnamefont {Grason}}, \bibinfo {author}
  {\bibfnamefont {M.~F.}\ \bibnamefont {Hagan}},\ and\ \bibinfo {author}
  {\bibfnamefont {Z.}~\bibnamefont {Dogic}},\ }\bibfield  {title} {\bibinfo
  {title} {Conformational switching of chiral colloidal rafts regulates
  raft--raft attractions and repulsions},\ }\href@noop {} {\bibfield  {journal}
  {\bibinfo  {journal} {Proceedings of the National Academy of Sciences}\
  }\textbf {\bibinfo {volume} {116}},\ \bibinfo {pages} {15792} (\bibinfo
  {year} {2019})}\BibitemShut {NoStop}%
\bibitem [{\citenamefont {Miller}\ \emph {et~al.}(2020)\citenamefont {Miller},
  \citenamefont {Hall}, \citenamefont {Robaszewski}, \citenamefont {Sharma},
  \citenamefont {Hagan}, \citenamefont {Grason},\ and\ \citenamefont
  {Dogic}}]{miller2020all}%
  \BibitemOpen
  \bibfield  {author} {\bibinfo {author} {\bibfnamefont {J.~M.}\ \bibnamefont
  {Miller}}, \bibinfo {author} {\bibfnamefont {D.}~\bibnamefont {Hall}},
  \bibinfo {author} {\bibfnamefont {J.}~\bibnamefont {Robaszewski}}, \bibinfo
  {author} {\bibfnamefont {P.}~\bibnamefont {Sharma}}, \bibinfo {author}
  {\bibfnamefont {M.~F.}\ \bibnamefont {Hagan}}, \bibinfo {author}
  {\bibfnamefont {G.~M.}\ \bibnamefont {Grason}},\ and\ \bibinfo {author}
  {\bibfnamefont {Z.}~\bibnamefont {Dogic}},\ }\bibfield  {title} {\bibinfo
  {title} {All twist and no bend makes raft edges splay: {S}pontaneous
  curvature of domain edges in colloidal membranes},\ }\href@noop {} {\bibfield
   {journal} {\bibinfo  {journal} {Science advances}\ }\textbf {\bibinfo
  {volume} {6}},\ \bibinfo {pages} {eaba2331} (\bibinfo {year}
  {2020})}\BibitemShut {NoStop}%
\bibitem [{\citenamefont {Dogic}\ and\ \citenamefont
  {Fraden}(2001)}]{dogic2001development}%
  \BibitemOpen
  \bibfield  {author} {\bibinfo {author} {\bibfnamefont {Z.}~\bibnamefont
  {Dogic}}\ and\ \bibinfo {author} {\bibfnamefont {S.}~\bibnamefont {Fraden}},\
  }\bibfield  {title} {\bibinfo {title} {Development of model colloidal liquid
  crystals and the kinetics of the isotropic--smectic transition},\ }\href@noop
  {} {\bibfield  {journal} {\bibinfo  {journal} {Philosophical Transactions of
  the Royal Society of London. Series A: Mathematical, Physical and Engineering
  Sciences}\ }\textbf {\bibinfo {volume} {359}},\ \bibinfo {pages} {997}
  (\bibinfo {year} {2001})}\BibitemShut {NoStop}%
\bibitem [{\citenamefont {Grelet}\ and\ \citenamefont
  {Fraden}(2003)}]{grelet2003origin}%
  \BibitemOpen
  \bibfield  {author} {\bibinfo {author} {\bibfnamefont {E.}~\bibnamefont
  {Grelet}}\ and\ \bibinfo {author} {\bibfnamefont {S.}~\bibnamefont
  {Fraden}},\ }\bibfield  {title} {\bibinfo {title} {What is the origin of
  chirality in the cholesteric phase of virus suspensions?},\ }\href@noop {}
  {\bibfield  {journal} {\bibinfo  {journal} {Physical Review Letters}\
  }\textbf {\bibinfo {volume} {90}},\ \bibinfo {pages} {198302} (\bibinfo
  {year} {2003})}\BibitemShut {NoStop}%
\bibitem [{\citenamefont {Barry}\ \emph {et~al.}(2009)\citenamefont {Barry},
  \citenamefont {Beller},\ and\ \citenamefont {Dogic}}]{barry2009model}%
  \BibitemOpen
  \bibfield  {author} {\bibinfo {author} {\bibfnamefont {E.}~\bibnamefont
  {Barry}}, \bibinfo {author} {\bibfnamefont {D.}~\bibnamefont {Beller}},\ and\
  \bibinfo {author} {\bibfnamefont {Z.}~\bibnamefont {Dogic}},\ }\bibfield
  {title} {\bibinfo {title} {A model liquid crystalline system based on rodlike
  viruses with variable chirality and persistence length},\ }\href@noop {}
  {\bibfield  {journal} {\bibinfo  {journal} {Soft Matter}\ }\textbf {\bibinfo
  {volume} {5}},\ \bibinfo {pages} {2563} (\bibinfo {year} {2009})}\BibitemShut
  {NoStop}%
\bibitem [{\citenamefont {Siavashpouri}\ \emph {et~al.}(2019)\citenamefont
  {Siavashpouri}, \citenamefont {Sharma}, \citenamefont {Fung}, \citenamefont
  {Hagan},\ and\ \citenamefont {Dogic}}]{siavashpouri2019structure}%
  \BibitemOpen
  \bibfield  {author} {\bibinfo {author} {\bibfnamefont {M.}~\bibnamefont
  {Siavashpouri}}, \bibinfo {author} {\bibfnamefont {P.}~\bibnamefont
  {Sharma}}, \bibinfo {author} {\bibfnamefont {J.}~\bibnamefont {Fung}},
  \bibinfo {author} {\bibfnamefont {M.~F.}\ \bibnamefont {Hagan}},\ and\
  \bibinfo {author} {\bibfnamefont {Z.}~\bibnamefont {Dogic}},\ }\bibfield
  {title} {\bibinfo {title} {Structure, dynamics and phase behavior of short
  rod inclusions dissolved in a colloidal membrane},\ }\href@noop {} {\bibfield
   {journal} {\bibinfo  {journal} {Soft matter}\ }\textbf {\bibinfo {volume}
  {15}},\ \bibinfo {pages} {7033} (\bibinfo {year} {2019})}\BibitemShut
  {NoStop}%
\bibitem [{\citenamefont {Dogic}(2003)}]{dogic2003surface}%
  \BibitemOpen
  \bibfield  {author} {\bibinfo {author} {\bibfnamefont {Z.}~\bibnamefont
  {Dogic}},\ }\bibfield  {title} {\bibinfo {title} {Surface freezing and a
  two-step pathway of the isotropic-smectic phase transition in colloidal
  rods},\ }\href@noop {} {\bibfield  {journal} {\bibinfo  {journal} {Physical
  review letters}\ }\textbf {\bibinfo {volume} {91}},\ \bibinfo {pages}
  {165701} (\bibinfo {year} {2003})}\BibitemShut {NoStop}%
\bibitem [{\citenamefont {Fomenko}\ and\ \citenamefont
  {Tuzhilin}(1991)}]{FomenkoTuzhilin1991}%
  \BibitemOpen
  \bibfield  {author} {\bibinfo {author} {\bibfnamefont {A.~T.}\ \bibnamefont
  {Fomenko}}\ and\ \bibinfo {author} {\bibfnamefont {A.~A.}\ \bibnamefont
  {Tuzhilin}},\ }\href@noop {} {\emph {\bibinfo {title} {Elements of the
  geometry and topology of minimal surfaces in three-dimensional space}}}\
  (\bibinfo  {publisher} {American Mathematical Society},\ \bibinfo {address}
  {Providence, RI},\ \bibinfo {year} {1991})\BibitemShut {NoStop}%
\bibitem [{\citenamefont {Canham}(1970)}]{canham1970minimum}%
  \BibitemOpen
  \bibfield  {author} {\bibinfo {author} {\bibfnamefont {P.~B.}\ \bibnamefont
  {Canham}},\ }\bibfield  {title} {\bibinfo {title} {The minimum energy of
  bending as a possible explanation of the biconcave shape of the human red
  blood cell},\ }\href@noop {} {\bibfield  {journal} {\bibinfo  {journal}
  {Journal of Theoretical Biology}\ }\textbf {\bibinfo {volume} {26}},\
  \bibinfo {pages} {61} (\bibinfo {year} {1970})}\BibitemShut {NoStop}%
\bibitem [{\citenamefont {Balchuas}\ \emph {et~al.}(2020)\citenamefont
  {Balchuas}, \citenamefont {Jia}, \citenamefont {Zakhary}, \citenamefont
  {Robaszewski}, \citenamefont {Gibaud}, , \citenamefont {Dogic}, \citenamefont
  {Pelcovits},\ and\ \citenamefont {Powers}}]{balchunas2020force}%
  \BibitemOpen
  \bibfield  {author} {\bibinfo {author} {\bibfnamefont {A.}~\bibnamefont
  {Balchuas}}, \bibinfo {author} {\bibfnamefont {L.~L.}\ \bibnamefont {Jia}},
  \bibinfo {author} {\bibfnamefont {M.~J.}\ \bibnamefont {Zakhary}}, \bibinfo
  {author} {\bibfnamefont {J.}~\bibnamefont {Robaszewski}}, \bibinfo {author}
  {\bibfnamefont {T.}~\bibnamefont {Gibaud}}, , \bibinfo {author}
  {\bibfnamefont {Z.}~\bibnamefont {Dogic}}, \bibinfo {author} {\bibfnamefont
  {R.~A.}\ \bibnamefont {Pelcovits}},\ and\ \bibinfo {author} {\bibfnamefont
  {T.~R.}\ \bibnamefont {Powers}},\ }\bibfield  {title} {\bibinfo {title}
  {Force-induced formation of twisted chiral ribbons},\ }\href@noop {}
  {\bibfield  {journal} {\bibinfo  {journal} {Phys. Rev. Lett.}\ }\textbf
  {\bibinfo {volume} {125}},\ \bibinfo {pages} {018002} (\bibinfo {year}
  {2020})}\BibitemShut {NoStop}%
\bibitem [{\citenamefont {Kl\'eman}(1983)}]{kleman}%
  \BibitemOpen
  \bibfield  {author} {\bibinfo {author} {\bibfnamefont {M.}~\bibnamefont
  {Kl\'eman}},\ }\href@noop {} {\emph {\bibinfo {title} {Points, Lines, and
  Walls}}}\ (\bibinfo  {publisher} {John Wiley \& Sons},\ \bibinfo {address}
  {Chichester},\ \bibinfo {year} {1983})\BibitemShut {NoStop}%
\bibitem [{\citenamefont {Balchunas}\ \emph
  {et~al.}(2019{\natexlab{b}})\citenamefont {Balchunas}, \citenamefont
  {Cabanas}, \citenamefont {Zakhary}, \citenamefont {Gibaud}, \citenamefont
  {Fraden}, \citenamefont {Sharma}, \citenamefont {Hagan},\ and\ \citenamefont
  {Dogic}}]{Balchunas_etal2019}%
  \BibitemOpen
  \bibfield  {author} {\bibinfo {author} {\bibfnamefont {A.~J.}\ \bibnamefont
  {Balchunas}}, \bibinfo {author} {\bibfnamefont {R.~A.}\ \bibnamefont
  {Cabanas}}, \bibinfo {author} {\bibfnamefont {M.~J.}\ \bibnamefont
  {Zakhary}}, \bibinfo {author} {\bibfnamefont {T.}~\bibnamefont {Gibaud}},
  \bibinfo {author} {\bibfnamefont {S.}~\bibnamefont {Fraden}}, \bibinfo
  {author} {\bibfnamefont {P.}~\bibnamefont {Sharma}}, \bibinfo {author}
  {\bibfnamefont {M.~F.}\ \bibnamefont {Hagan}},\ and\ \bibinfo {author}
  {\bibfnamefont {Z.}~\bibnamefont {Dogic}},\ }\bibfield  {title} {\bibinfo
  {title} {Equation of state of colloidal membranes},\ }\href@noop {}
  {\bibfield  {journal} {\bibinfo  {journal} {Soft Matter}\ }\textbf {\bibinfo
  {volume} {15}},\ \bibinfo {pages} {6791} (\bibinfo {year}
  {2019}{\natexlab{b}})}\BibitemShut {NoStop}%
\bibitem [{\citenamefont {Zakhary}\ \emph {et~al.}(2014)\citenamefont
  {Zakhary}, \citenamefont {Gibaud}, \citenamefont {Kaplan}, \citenamefont
  {Barry}, \citenamefont {Oldenbourg}, \citenamefont {Meyer},\ and\
  \citenamefont {Dogic}}]{zakhary2014imprintable}%
  \BibitemOpen
  \bibfield  {author} {\bibinfo {author} {\bibfnamefont {M.~J.}\ \bibnamefont
  {Zakhary}}, \bibinfo {author} {\bibfnamefont {T.}~\bibnamefont {Gibaud}},
  \bibinfo {author} {\bibfnamefont {C.~N.}\ \bibnamefont {Kaplan}}, \bibinfo
  {author} {\bibfnamefont {E.}~\bibnamefont {Barry}}, \bibinfo {author}
  {\bibfnamefont {R.}~\bibnamefont {Oldenbourg}}, \bibinfo {author}
  {\bibfnamefont {R.~B.}\ \bibnamefont {Meyer}},\ and\ \bibinfo {author}
  {\bibfnamefont {Z.}~\bibnamefont {Dogic}},\ }\bibfield  {title} {\bibinfo
  {title} {Imprintable membranes from incomplete chiral coalescence},\
  }\href@noop {} {\bibfield  {journal} {\bibinfo  {journal} {Nature
  Communications}\ }\textbf {\bibinfo {volume} {5}},\ \bibinfo {pages} {3063}
  (\bibinfo {year} {2014})}\BibitemShut {NoStop}%
\bibitem [{\citenamefont {Stanley}\ and\ \citenamefont
  {Strey}(2003)}]{PEG_osmotic_pressure}%
  \BibitemOpen
  \bibfield  {author} {\bibinfo {author} {\bibfnamefont {C.~B.}\ \bibnamefont
  {Stanley}}\ and\ \bibinfo {author} {\bibfnamefont {H.~H.}\ \bibnamefont
  {Strey}},\ }\bibfield  {title} {\bibinfo {title} {Measuring osmotic pressure
  of poly(ethylene glycol) solutions by sedimentation equilibrium
  ultracentrifugation},\ }\href@noop {} {\bibfield  {journal} {\bibinfo
  {journal} {Macromolecules}\ }\textbf {\bibinfo {volume} {36}},\ \bibinfo
  {pages} {6888} (\bibinfo {year} {2003})}\BibitemShut {NoStop}%
\bibitem [{\citenamefont {Harbich}\ \emph {et~al.}(1978)\citenamefont
  {Harbich}, \citenamefont {Servuss},\ and\ \citenamefont
  {Helfrich}}]{Harbich_etal1978}%
  \BibitemOpen
  \bibfield  {author} {\bibinfo {author} {\bibfnamefont {W.}~\bibnamefont
  {Harbich}}, \bibinfo {author} {\bibfnamefont {R.}~\bibnamefont {Servuss}},\
  and\ \bibinfo {author} {\bibfnamefont {W.}~\bibnamefont {Helfrich}},\
  }\bibfield  {title} {\bibinfo {title} {Passages in lecithin-water systems},\
  }\href@noop {} {\bibfield  {journal} {\bibinfo  {journal} {Z. Naturforsch.}\
  }\textbf {\bibinfo {volume} {{\bf 33a}}},\ \bibinfo {pages} {1013} (\bibinfo
  {year} {1978})}\BibitemShut {NoStop}%
\bibitem [{\citenamefont {Helfrich}(1981)}]{Helfrich1981}%
  \BibitemOpen
  \bibfield  {author} {\bibinfo {author} {\bibfnamefont {W.}~\bibnamefont
  {Helfrich}},\ }in\ \href@noop {} {\emph {\bibinfo {booktitle} {Physics of
  Defects (Les Houches, Session XXXV, 1980}}},\ \bibinfo {editor} {edited by\
  \bibinfo {editor} {\bibfnamefont {R.}~\bibnamefont {Balian}}, \bibinfo
  {editor} {\bibfnamefont {M.}~\bibnamefont {Kl\'eman}},\ and\ \bibinfo
  {editor} {\bibfnamefont {J.~P.}\ \bibnamefont {Poirier}}}\ (\bibinfo
  {publisher} {North Holland},\ \bibinfo {address} {Amsterdam},\ \bibinfo
  {year} {1981})\ p.\ \bibinfo {pages} {715}\BibitemShut {NoStop}%
\bibitem [{\citenamefont {Terzi}\ \emph {et~al.}(2019)\citenamefont {Terzi},
  \citenamefont {Ergr\"uder},\ and\ \citenamefont
  {Deserno}}]{TerziErgruderDeserno2019}%
  \BibitemOpen
  \bibfield  {author} {\bibinfo {author} {\bibfnamefont {M.~M.}\ \bibnamefont
  {Terzi}}, \bibinfo {author} {\bibfnamefont {M.~F.}\ \bibnamefont
  {Ergr\"uder}},\ and\ \bibinfo {author} {\bibfnamefont {M.}~\bibnamefont
  {Deserno}},\ }\bibfield  {title} {\bibinfo {title} {A consistent quadratic
  curvature-tilt theory for fluid membranes},\ }\href@noop {} {\bibfield
  {journal} {\bibinfo  {journal} {J. Chem. Phys.}\ }\textbf {\bibinfo {volume}
  {151}},\ \bibinfo {pages} {164108} (\bibinfo {year} {2019})}\BibitemShut
  {NoStop}%
\bibitem [{\citenamefont {Johannes}\ \emph {et~al.}(2014)\citenamefont
  {Johannes}, \citenamefont {Wunder},\ and\ \citenamefont
  {Bassereau}}]{endocytosis1}%
  \BibitemOpen
  \bibfield  {author} {\bibinfo {author} {\bibfnamefont {L.}~\bibnamefont
  {Johannes}}, \bibinfo {author} {\bibfnamefont {C.}~\bibnamefont {Wunder}},\
  and\ \bibinfo {author} {\bibfnamefont {P.}~\bibnamefont {Bassereau}},\
  }\bibfield  {title} {\bibinfo {title} {Bending "on the rocks"--a cocktail of
  biophysical modules to build endocytic pathways},\ }\href@noop {} {\bibfield
  {journal} {\bibinfo  {journal} {Cold Spring Harbor perspectives in biology}\
  }\textbf {\bibinfo {volume} {6}},\ \bibinfo {pages} {a016741} (\bibinfo
  {year} {2014})}\BibitemShut {NoStop}%
\bibitem [{\citenamefont {McMahon}\ and\ \citenamefont
  {Boucrot}(2011)}]{endocytosis2}%
  \BibitemOpen
  \bibfield  {author} {\bibinfo {author} {\bibfnamefont {H.~T.}\ \bibnamefont
  {McMahon}}\ and\ \bibinfo {author} {\bibfnamefont {E.}~\bibnamefont
  {Boucrot}},\ }\bibfield  {title} {\bibinfo {title} {Molecular mechanism and
  physiological functions of clathrin-mediated endocytosis},\ }\href@noop {}
  {\bibfield  {journal} {\bibinfo  {journal} {Nature Reviews Molecular Cell
  Biology}\ }\textbf {\bibinfo {volume} {12}},\ \bibinfo {pages} {517}
  (\bibinfo {year} {2011})}\BibitemShut {NoStop}%
\bibitem [{\citenamefont {Südhof}\ and\ \citenamefont
  {Rizo}(2011)}]{exocytosis1}%
  \BibitemOpen
  \bibfield  {author} {\bibinfo {author} {\bibfnamefont {T.~C.}\ \bibnamefont
  {Südhof}}\ and\ \bibinfo {author} {\bibfnamefont {J.}~\bibnamefont {Rizo}},\
  }\bibfield  {title} {\bibinfo {title} {Synaptic vesicle exocytosis},\
  }\href@noop {} {\bibfield  {journal} {\bibinfo  {journal} {Cold Spring Harbor
  perspectives in biology}\ }\textbf {\bibinfo {volume} {3}},\ \bibinfo {pages}
  {a005637} (\bibinfo {year} {2011})}\BibitemShut {NoStop}%
\bibitem [{\citenamefont {Westermann}(2010)}]{fission1}%
  \BibitemOpen
  \bibfield  {author} {\bibinfo {author} {\bibfnamefont {B.}~\bibnamefont
  {Westermann}},\ }\bibfield  {title} {\bibinfo {title} {Mitochondrial fusion
  and fission in cell life and death},\ }\href@noop {} {\bibfield  {journal}
  {\bibinfo  {journal} {Nature Reviews Molecular Cell Biology}\ }\textbf
  {\bibinfo {volume} {11}},\ \bibinfo {pages} {872} (\bibinfo {year}
  {2010})}\BibitemShut {NoStop}%
\bibitem [{\citenamefont {Rangamani}\ \emph {et~al.}(2011)\citenamefont
  {Rangamani}, \citenamefont {Fardin}, \citenamefont {Xiong}, \citenamefont
  {Lipshtat}, \citenamefont {Rossier}, \citenamefont {Sheetz},\ and\
  \citenamefont {Iyengar}}]{cellmotility1}%
  \BibitemOpen
  \bibfield  {author} {\bibinfo {author} {\bibfnamefont {P.}~\bibnamefont
  {Rangamani}}, \bibinfo {author} {\bibfnamefont {M.-A.}\ \bibnamefont
  {Fardin}}, \bibinfo {author} {\bibfnamefont {Y.}~\bibnamefont {Xiong}},
  \bibinfo {author} {\bibfnamefont {A.}~\bibnamefont {Lipshtat}}, \bibinfo
  {author} {\bibfnamefont {O.}~\bibnamefont {Rossier}}, \bibinfo {author}
  {\bibfnamefont {M.~P.}\ \bibnamefont {Sheetz}},\ and\ \bibinfo {author}
  {\bibfnamefont {R.}~\bibnamefont {Iyengar}},\ }\bibfield  {title} {\bibinfo
  {title} {Signaling network triggers and membrane physical properties control
  the actin cytoskeleton-driven isotropic phase of cell spreading},\
  }\href@noop {} {\bibfield  {journal} {\bibinfo  {journal} {Biophysical
  journal}\ }\textbf {\bibinfo {volume} {100}},\ \bibinfo {pages} {845}
  (\bibinfo {year} {2011})}\BibitemShut {NoStop}%
\bibitem [{\citenamefont {Pontes}\ \emph {et~al.}(2013)\citenamefont {Pontes},
  \citenamefont {Ayala}, \citenamefont {Fonseca}, \citenamefont {Romao},
  \citenamefont {Amaral},\ and\ \citenamefont {Salgado}}]{cellfunction1}%
  \BibitemOpen
  \bibfield  {author} {\bibinfo {author} {\bibfnamefont {B.}~\bibnamefont
  {Pontes}}, \bibinfo {author} {\bibfnamefont {Y.}~\bibnamefont {Ayala}},
  \bibinfo {author} {\bibfnamefont {A.}~\bibnamefont {Fonseca}}, \bibinfo
  {author} {\bibfnamefont {L.}~\bibnamefont {Romao}}, \bibinfo {author}
  {\bibfnamefont {R.}~\bibnamefont {Amaral}},\ and\ \bibinfo {author}
  {\bibfnamefont {L.~e.~a.}\ \bibnamefont {Salgado}},\ }\bibfield  {title}
  {\bibinfo {title} {Membrane elastic properties and cell function},\
  }\href@noop {} {\bibfield  {journal} {\bibinfo  {journal} {PLoS ONE}\
  }\textbf {\bibinfo {volume} {8}},\ \bibinfo {pages} {e67708} (\bibinfo {year}
  {2013})}\BibitemShut {NoStop}%
\bibitem [{\citenamefont {Hu}\ \emph {et~al.}(2012)\citenamefont {Hu},
  \citenamefont {Briguglio},\ and\ \citenamefont {Deserno}}]{deserno1}%
  \BibitemOpen
  \bibfield  {author} {\bibinfo {author} {\bibfnamefont {M.}~\bibnamefont
  {Hu}}, \bibinfo {author} {\bibfnamefont {J.}~\bibnamefont {Briguglio}},\ and\
  \bibinfo {author} {\bibfnamefont {M.}~\bibnamefont {Deserno}},\ }\bibfield
  {title} {\bibinfo {title} {Determining the {G}aussian curvature modulus of
  lipid membranes in simulations},\ }\href@noop {} {\bibfield  {journal}
  {\bibinfo  {journal} {Biophysical Journal}\ }\textbf {\bibinfo {volume}
  {102}},\ \bibinfo {pages} {1403} (\bibinfo {year} {2012})}\BibitemShut
  {NoStop}%
\bibitem [{\citenamefont {Siegel}\ and\ \citenamefont
  {Kozlov}(2004)}]{kozlov2}%
  \BibitemOpen
  \bibfield  {author} {\bibinfo {author} {\bibfnamefont {D.~P.}\ \bibnamefont
  {Siegel}}\ and\ \bibinfo {author} {\bibfnamefont {M.~M.}\ \bibnamefont
  {Kozlov}},\ }\bibfield  {title} {\bibinfo {title} {The {G}aussian curvature
  elastic modulus of {N}-monomethylated dioleoylphosphatidylethanolamine:
  {R}elevance to membrane fusion and lipid phase behavior},\ }\href@noop {}
  {\bibfield  {journal} {\bibinfo  {journal} {Biophysical journal}\ }\textbf
  {\bibinfo {volume} {87}},\ \bibinfo {pages} {366} (\bibinfo {year}
  {2004})}\BibitemShut {NoStop}%
\bibitem [{\citenamefont {Fonda}\ \emph {et~al.}(2020)\citenamefont {Fonda},
  \citenamefont {Al-Izzi}, \citenamefont {Giomi},\ and\ \citenamefont
  {Turner}}]{turner1}%
  \BibitemOpen
  \bibfield  {author} {\bibinfo {author} {\bibfnamefont {P.}~\bibnamefont
  {Fonda}}, \bibinfo {author} {\bibfnamefont {S.~C.}\ \bibnamefont {Al-Izzi}},
  \bibinfo {author} {\bibfnamefont {L.}~\bibnamefont {Giomi}},\ and\ \bibinfo
  {author} {\bibfnamefont {M.~S.}\ \bibnamefont {Turner}},\ }\bibfield  {title}
  {\bibinfo {title} {Measuring {G}aussian rigidity using curved substrates},\
  }\href@noop {} {\bibfield  {journal} {\bibinfo  {journal} {Phys. Rev. Lett.}\
  }\textbf {\bibinfo {volume} {125}},\ \bibinfo {pages} {188002} (\bibinfo
  {year} {2020})}\BibitemShut {NoStop}%
\bibitem [{\citenamefont {Baumgart}\ \emph {et~al.}(2005)\citenamefont
  {Baumgart}, \citenamefont {Das}, \citenamefont {Webb},\ and\ \citenamefont
  {Jenkins}}]{baumgart1}%
  \BibitemOpen
  \bibfield  {author} {\bibinfo {author} {\bibfnamefont {T.}~\bibnamefont
  {Baumgart}}, \bibinfo {author} {\bibfnamefont {S.}~\bibnamefont {Das}},
  \bibinfo {author} {\bibfnamefont {W.~W.}\ \bibnamefont {Webb}},\ and\
  \bibinfo {author} {\bibfnamefont {J.~T.}\ \bibnamefont {Jenkins}},\
  }\bibfield  {title} {\bibinfo {title} {Membrane elasticity in giant vesicles
  with fluid phase coexistence},\ }\href@noop {} {\bibfield  {journal}
  {\bibinfo  {journal} {Biophysical journal}\ }\textbf {\bibinfo {volume}
  {89}},\ \bibinfo {pages} {1067} (\bibinfo {year} {2005})}\BibitemShut
  {NoStop}%
\bibitem [{\citenamefont {Semrau}\ \emph {et~al.}(2008)\citenamefont {Semrau},
  \citenamefont {Idema}, \citenamefont {Holtzer}, \citenamefont {Schmidt},\
  and\ \citenamefont {Storm}}]{gauss1}%
  \BibitemOpen
  \bibfield  {author} {\bibinfo {author} {\bibfnamefont {S.}~\bibnamefont
  {Semrau}}, \bibinfo {author} {\bibfnamefont {T.}~\bibnamefont {Idema}},
  \bibinfo {author} {\bibfnamefont {L.}~\bibnamefont {Holtzer}}, \bibinfo
  {author} {\bibfnamefont {T.}~\bibnamefont {Schmidt}},\ and\ \bibinfo {author}
  {\bibfnamefont {C.}~\bibnamefont {Storm}},\ }\bibfield  {title} {\bibinfo
  {title} {Accurate determination of elastic parameters for multicomponent
  membranes},\ }\href@noop {} {\bibfield  {journal} {\bibinfo  {journal} {Phys.
  Rev. Lett.}\ }\textbf {\bibinfo {volume} {100}},\ \bibinfo {pages} {088101}
  (\bibinfo {year} {2008})}\BibitemShut {NoStop}%
\bibitem [{\citenamefont {Pusey}\ and\ \citenamefont {van
  Megen}(1986)}]{colloids1}%
  \BibitemOpen
  \bibfield  {author} {\bibinfo {author} {\bibfnamefont {P.~N.}\ \bibnamefont
  {Pusey}}\ and\ \bibinfo {author} {\bibfnamefont {W.}~\bibnamefont {van
  Megen}},\ }\bibfield  {title} {\bibinfo {title} {Phase behaviour of
  concentrated suspensions of nearly hard colloidal spheres},\ }\href@noop {}
  {\bibfield  {journal} {\bibinfo  {journal} {Nature}\ }\textbf {\bibinfo
  {volume} {320}},\ \bibinfo {pages} {340} (\bibinfo {year}
  {1986})}\BibitemShut {NoStop}%
\bibitem [{\citenamefont {van Blaaderen}\ and\ \citenamefont
  {Wiltzius}(1995)}]{van1995real}%
  \BibitemOpen
  \bibfield  {author} {\bibinfo {author} {\bibfnamefont {A.}~\bibnamefont {van
  Blaaderen}}\ and\ \bibinfo {author} {\bibfnamefont {P.}~\bibnamefont
  {Wiltzius}},\ }\bibfield  {title} {\bibinfo {title} {Real-space structure of
  colloidal hard-sphere glasses},\ }\href@noop {} {\bibfield  {journal}
  {\bibinfo  {journal} {Science}\ }\textbf {\bibinfo {volume} {270}},\ \bibinfo
  {pages} {1177} (\bibinfo {year} {1995})}\BibitemShut {NoStop}%
\bibitem [{\citenamefont {Schall}\ \emph {et~al.}(2006)\citenamefont {Schall},
  \citenamefont {Cohen}, \citenamefont {Weitz},\ and\ \citenamefont
  {Spaepen}}]{schall2006visualizing}%
  \BibitemOpen
  \bibfield  {author} {\bibinfo {author} {\bibfnamefont {P.}~\bibnamefont
  {Schall}}, \bibinfo {author} {\bibfnamefont {I.}~\bibnamefont {Cohen}},
  \bibinfo {author} {\bibfnamefont {D.~A.}\ \bibnamefont {Weitz}},\ and\
  \bibinfo {author} {\bibfnamefont {F.}~\bibnamefont {Spaepen}},\ }\bibfield
  {title} {\bibinfo {title} {Visualizing dislocation nucleation by indenting
  colloidal crystals},\ }\href@noop {} {\bibfield  {journal} {\bibinfo
  {journal} {Nature}\ }\textbf {\bibinfo {volume} {440}},\ \bibinfo {pages}
  {319} (\bibinfo {year} {2006})}\BibitemShut {NoStop}%
\bibitem [{\citenamefont {He}\ \emph {et~al.}(2020)\citenamefont {He},
  \citenamefont {Gales}, \citenamefont {Ducrot}, \citenamefont {Gong},
  \citenamefont {Yi}, \citenamefont {Sacanna},\ and\ \citenamefont
  {Pine}}]{he2020colloidal}%
  \BibitemOpen
  \bibfield  {author} {\bibinfo {author} {\bibfnamefont {M.}~\bibnamefont
  {He}}, \bibinfo {author} {\bibfnamefont {J.~P.}\ \bibnamefont {Gales}},
  \bibinfo {author} {\bibfnamefont {{\'E}.}~\bibnamefont {Ducrot}}, \bibinfo
  {author} {\bibfnamefont {Z.}~\bibnamefont {Gong}}, \bibinfo {author}
  {\bibfnamefont {G.-R.}\ \bibnamefont {Yi}}, \bibinfo {author} {\bibfnamefont
  {S.}~\bibnamefont {Sacanna}},\ and\ \bibinfo {author} {\bibfnamefont {D.~J.}\
  \bibnamefont {Pine}},\ }\bibfield  {title} {\bibinfo {title} {Colloidal
  diamond},\ }\href@noop {} {\bibfield  {journal} {\bibinfo  {journal}
  {Nature}\ }\textbf {\bibinfo {volume} {585}},\ \bibinfo {pages} {524}
  (\bibinfo {year} {2020})}\BibitemShut {NoStop}%
\bibitem [{\citenamefont {Demurtas}\ \emph {et~al.}(2015)\citenamefont
  {Demurtas}, \citenamefont {Guichard}, \citenamefont {Martiel}, \citenamefont
  {Mezzenga}, \citenamefont {H{\'e}bert},\ and\ \citenamefont
  {Sagalowicz}}]{Demurtas2015}%
  \BibitemOpen
  \bibfield  {author} {\bibinfo {author} {\bibfnamefont {D.}~\bibnamefont
  {Demurtas}}, \bibinfo {author} {\bibfnamefont {P.}~\bibnamefont {Guichard}},
  \bibinfo {author} {\bibfnamefont {I.}~\bibnamefont {Martiel}}, \bibinfo
  {author} {\bibfnamefont {R.}~\bibnamefont {Mezzenga}}, \bibinfo {author}
  {\bibfnamefont {C.}~\bibnamefont {H{\'e}bert}},\ and\ \bibinfo {author}
  {\bibfnamefont {L.}~\bibnamefont {Sagalowicz}},\ }\bibfield  {title}
  {\bibinfo {title} {Direct visualization of dispersed lipid bicontinuous cubic
  phases by cryo-electron tomography},\ }\href@noop {} {\bibfield  {journal}
  {\bibinfo  {journal} {Nature Communications}\ }\textbf {\bibinfo {volume}
  {6}},\ \bibinfo {pages} {8915} (\bibinfo {year} {2015})}\BibitemShut
  {NoStop}%
\bibitem [{\citenamefont {Barriga}\ \emph {et~al.}(2019)\citenamefont
  {Barriga}, \citenamefont {Holme},\ and\ \citenamefont
  {Stevens}}]{Cubosome_Barriga}%
  \BibitemOpen
  \bibfield  {author} {\bibinfo {author} {\bibfnamefont {H.~M.~G.}\
  \bibnamefont {Barriga}}, \bibinfo {author} {\bibfnamefont {M.~N.}\
  \bibnamefont {Holme}},\ and\ \bibinfo {author} {\bibfnamefont {M.~M.}\
  \bibnamefont {Stevens}},\ }\bibfield  {title} {\bibinfo {title} {Cubosomes:
  {T}he next generation of smart lipid nanoparticles?},\ }\href@noop {}
  {\bibfield  {journal} {\bibinfo  {journal} {Angewandte Chemie International
  Edition}\ }\textbf {\bibinfo {volume} {58}},\ \bibinfo {pages} {2958}
  (\bibinfo {year} {2019})}\BibitemShut {NoStop}%
\bibitem [{\citenamefont {Gazeau}\ \emph {et~al.}(1989)\citenamefont {Gazeau},
  \citenamefont {Bellocq}, \citenamefont {Roux},\ and\ \citenamefont
  {Zemb}}]{Gazeau_1989}%
  \BibitemOpen
  \bibfield  {author} {\bibinfo {author} {\bibfnamefont {D.}~\bibnamefont
  {Gazeau}}, \bibinfo {author} {\bibfnamefont {A.~M.}\ \bibnamefont {Bellocq}},
  \bibinfo {author} {\bibfnamefont {D.}~\bibnamefont {Roux}},\ and\ \bibinfo
  {author} {\bibfnamefont {T.}~\bibnamefont {Zemb}},\ }\bibfield  {title}
  {\bibinfo {title} {Experimental evidence for random surface structures in
  dilute surfactant solutions},\ }\href@noop {} {\bibfield  {journal} {\bibinfo
   {journal} {Europhysics Letters ({EPL})}\ }\textbf {\bibinfo {volume} {9}},\
  \bibinfo {pages} {447} (\bibinfo {year} {1989})}\BibitemShut {NoStop}%
\bibitem [{\citenamefont {Cates}\ \emph {et~al.}(1988)\citenamefont {Cates},
  \citenamefont {Roux}, \citenamefont {Andelman}, \citenamefont {Milner},\ and\
  \citenamefont {Safran}}]{Cates_1988}%
  \BibitemOpen
  \bibfield  {author} {\bibinfo {author} {\bibfnamefont {M.~E.}\ \bibnamefont
  {Cates}}, \bibinfo {author} {\bibfnamefont {D.}~\bibnamefont {Roux}},
  \bibinfo {author} {\bibfnamefont {D.}~\bibnamefont {Andelman}}, \bibinfo
  {author} {\bibfnamefont {S.~T.}\ \bibnamefont {Milner}},\ and\ \bibinfo
  {author} {\bibfnamefont {S.~A.}\ \bibnamefont {Safran}},\ }\bibfield  {title}
  {\bibinfo {title} {Random surface model for the $l_3$-phase of dilute
  surfactant solutions},\ }\href@noop {} {\bibfield  {journal} {\bibinfo
  {journal} {Europhysics Letters ({EPL})}\ }\textbf {\bibinfo {volume} {5}},\
  \bibinfo {pages} {733} (\bibinfo {year} {1988})}\BibitemShut {NoStop}%
\bibitem [{\citenamefont {Keim}\ \emph {et~al.}(2006)\citenamefont {Keim},
  \citenamefont {M\o{}ller}, \citenamefont {Zhang},\ and\ \citenamefont
  {Nagel}}]{singularity1}%
  \BibitemOpen
  \bibfield  {author} {\bibinfo {author} {\bibfnamefont {N.~C.}\ \bibnamefont
  {Keim}}, \bibinfo {author} {\bibfnamefont {P.}~\bibnamefont {M\o{}ller}},
  \bibinfo {author} {\bibfnamefont {W.~W.}\ \bibnamefont {Zhang}},\ and\
  \bibinfo {author} {\bibfnamefont {S.~R.}\ \bibnamefont {Nagel}},\ }\bibfield
  {title} {\bibinfo {title} {Breakup of air bubbles in water: {M}emory and
  breakdown of cylindrical symmetry},\ }\href@noop {} {\bibfield  {journal}
  {\bibinfo  {journal} {Phys. Rev. Lett.}\ }\textbf {\bibinfo {volume} {97}},\
  \bibinfo {pages} {144503} (\bibinfo {year} {2006})}\BibitemShut {NoStop}%
\bibitem [{\citenamefont {Xiong}\ \emph {et~al.}(2020)\citenamefont {Xiong},
  \citenamefont {Cao}, \citenamefont {Cooper}, \citenamefont {Rappel},
  \citenamefont {Hasty},\ and\ \citenamefont {Tsimring}}]{singularity2}%
  \BibitemOpen
  \bibfield  {author} {\bibinfo {author} {\bibfnamefont {L.}~\bibnamefont
  {Xiong}}, \bibinfo {author} {\bibfnamefont {Y.}~\bibnamefont {Cao}}, \bibinfo
  {author} {\bibfnamefont {R.}~\bibnamefont {Cooper}}, \bibinfo {author}
  {\bibfnamefont {W.}~\bibnamefont {Rappel}}, \bibinfo {author} {\bibfnamefont
  {J.}~\bibnamefont {Hasty}},\ and\ \bibinfo {author} {\bibfnamefont
  {L.}~\bibnamefont {Tsimring}},\ }\bibfield  {title} {\bibinfo {title}
  {Flower-like patterns in multi-species bacterial colonies},\ }\href@noop {}
  {\bibfield  {journal} {\bibinfo  {journal} {eLife}\ }\textbf {\bibinfo
  {volume} {9}},\ \bibinfo {pages} {e48885} (\bibinfo {year}
  {2020})}\BibitemShut {NoStop}%
\bibitem [{\citenamefont {Huang}\ \emph {et~al.}(2010)\citenamefont {Huang},
  \citenamefont {Davidovitch}, \citenamefont {Santangelo}, \citenamefont
  {Russell},\ and\ \citenamefont {Menon}}]{Menon1}%
  \BibitemOpen
  \bibfield  {author} {\bibinfo {author} {\bibfnamefont {J.}~\bibnamefont
  {Huang}}, \bibinfo {author} {\bibfnamefont {B.}~\bibnamefont {Davidovitch}},
  \bibinfo {author} {\bibfnamefont {C.~D.}\ \bibnamefont {Santangelo}},
  \bibinfo {author} {\bibfnamefont {T.~P.}\ \bibnamefont {Russell}},\ and\
  \bibinfo {author} {\bibfnamefont {N.}~\bibnamefont {Menon}},\ }\bibfield
  {title} {\bibinfo {title} {Smooth cascade of wrinkles at the edge of a
  floating elastic film},\ }\href@noop {} {\bibfield  {journal} {\bibinfo
  {journal} {Phys. Rev. Lett.}\ }\textbf {\bibinfo {volume} {105}},\ \bibinfo
  {pages} {038302} (\bibinfo {year} {2010})}\BibitemShut {NoStop}%
\bibitem [{\citenamefont {Sambrook}\ \emph {et~al.}(1989)\citenamefont
  {Sambrook}, \citenamefont {Fritsch},\ and\ \citenamefont
  {Maniatis}}]{Sambrook}%
  \BibitemOpen
  \bibfield  {author} {\bibinfo {author} {\bibfnamefont {J.}~\bibnamefont
  {Sambrook}}, \bibinfo {author} {\bibfnamefont {E.~F.}\ \bibnamefont
  {Fritsch}},\ and\ \bibinfo {author} {\bibfnamefont {T.}~\bibnamefont
  {Maniatis}},\ }\href@noop {} {\emph {\bibinfo {title} {Molecular cloning :
  {A} laboratory manual}}}\ (\bibinfo  {publisher} {Cold Spring Harbor
  Laboratory Press},\ \bibinfo {year} {1989})\BibitemShut {NoStop}%
\end{thebibliography}%

\end{document}